**Computational Agent-based Models in Opinion Dynamics:**

**A Survey on Social Simulations and Empirical Studies**

**Authors:** Yun-Shiuan Chuang[1,2], Timothy T. Rogers[1]

**Affiliations:**

[1]Department of Psychology, University of Wisconsin-Madison, Madison, USA 53706.

[2] Department of Computer Science, University of Wisconsin-Madison, Madison, USA 53706.

## Abstract

Understanding how an individual changes its attitude, belief, and opinion due to other people's social influences is vital because of its wide implications. A core methodology that is used to study the change of attitude under social influences is agent-based model (ABM). The goal of this review paper is to compare and contrast existing ABMs, which I classify into two families, the deductive ABMs and the inductive ABMs. The former subsumes social simulation studies, and the latter involves human experiments. To facilitate the comparison between ABMs of different formulations, I propose a general unified formulation, in which all ABMs can be viewed as special cases. In addition, I show the connections between deductive ABMs and inductive ABMs, and point out their strengths and limitations. At the end of the paper, I identify underexplored areas and suggest future research directions.

**Correspondence to:**

**Yun-Shiuan Chuang (ychuang26@wisc.edu), Timothy T. Rogers (ttrogers@wisc.edu)**

## Introduction

Understanding how an individual changes its attitude, belief, and opinion due to other people's social influences is vital because of its wide implications. For example, undesired societal phenomena like opinion bipolarization and opinion extremization is often a product of social influences among people (Flache et al., 2017). Another implication is the mitigation of misinformation, as misinformation can spread rapidly thanks to social influences (Budak et al., 2011). Aside from political opinions, the study of social influence on attitude has applications in marketing as well, where the adoption of a new product is impacted by social influences. A reliable understanding of how people change their attitude due to social influences allow forecasting the future (e.g., whether Twitter users will become bipolarized on a specific topic), and devising intervention programs (e.g., who should we target at and what message should we deliver to mitigate bipolarization in a population).

A core methodology that is used to study the change of attitude under social influences is agent-based model (ABM). In this context of studying attitude formation, ABMs are micro-level mathematical models that define how an agent (in this context, an agent usually represents a person) should update its attitude toward a specific topic upon receiving other people's opinions. I limit the scope of the review with the following criteria based on psychological understandings of attitude. First, the attitude variable in the ABM should be continuous with an origin. That is, one can define a positive attitude versus a negative attitude (e.g., in favor of a policy versus against a policy), and the magnitude of the value describe the intensity of the attitude (e.g., how much a person is in favor of a policy). Under this criterion, diffusion models in the literature of influence maximization or epidemiology are not included because an agent can only be uninfluenced/uninfected or influenced/infected (Banerjee et al., 2020; Li et al., 2018). Second, the change of attitude should be bidirectional. That is, if a person changes his attitude from negative to positive, he should be allowed to change back. With these criterion, diffusion models in the literature of influence limitation are not included because an agent cannot change its attitude once adopted one (Budak et al., 2011).

The goal of this review paper is to compare and contrast existing ABMs that describe how an individual updates his attitude upon receiving information provided by other people. This scope covers work from two primary approaches - the *deductive ABMs* and the *inductive ABMs*. Deductive ABMs refer to the models that are proposed based on deduction from some "first principles". These first principles are often grounded on psychological theories about how humans respond to social information. On the other hand, inductive ABMs refer to the approach where models are induced from empirical data of human experiments, in which the human subjects' attitudes are measured under social influences. To facilitate the comparison between ABMs of different formulations, I propose a general unified formulation, in which all ABMs can be viewed as special cases. In addition, I show the connections between deductive ABMs and inductive ABMs, and point out their strength and constraints. At the end of the paper, I identify underexplored areas and suggest future research directions.

## Deductive ABMs: Social Simulation

The study of social simulation is an attempt to explain the macro-level societal phenomena using ABMs. Within this field, the macro-level distribution of people's attitudes (e.g., bipolarization, echo chamber, extremism, consensus formation) is a particular phenomenon that scholars aim to simulate. For any given ABM, some macro-level attitude distributions can be simulated by the interactions between agents while some other may not. For example, an ABM may be able to simulate consensus formation but fail to generate bipolarization. The idea of explaining macro-level attitude distributions with ABMs has attracted attention from a wide variety of disciplines, including sociology (Banisch & Olbrich, 2019; Flache et al., 2017), cognitive science (Lorenz et al., 2021; Anderson, 1971; Hunter et al., 1984), communication science (Centola et al., 2018), statistics (DeGroot, 1974), philosophy (Wagner, 1978), and even physics (Baumann et al., 2020; Weisbuch et al., 2005), among many. Scholars from different disciplines have proposed various ABMs based on different theories. Among these ABMs, three major approaches emerge, 1) opinion dynamics, 2) Bayesian agents, and 3) physics-like models. First, the goal of the *opinion dynamics* research is to simulate macro-level attitude distributions (e.g., bipolarization) using ABMs inspired by psychological theories (e.g., the theory of confirmation bias predicts that people tend to be more receptive to information similar to their existing attitude), but the exact formulations of the ABMs are rarely verified empirically (Flache et al., 2017). On the other hand, Bayesian ABMs are proposed with the intent to demonstrate whether specific macro-level attitude distributions (e.g., echo chambers) can occur even with perfectly rational learning agents without any cognitive bias (Madsen et al., 2018). Finally, there are physics-like ABMs proposed by physicist in an attempt to simulate opinion formation (Baumann et al., 2020; Holme & Newman, 2006). These ABMs are rarely grounded on psychological theories, but nonetheless succeed in simulating macro-level attitude distribution observed on social media (Baumann et al., 2020).

Although the exact formulations of the ABMs vary, e.g., weighted averaging agents (DeGroot, 1974), Bayesian agents (Madsen et al., 2018), differential equations (Baumann et al., 2020), most models can be decomposed into components that are present across different ABMs. In this review paper, I identify four components in ABMs: 1) assimilation force, 2) reinforcement force, 3) similarity bias, and 4) repulsion force. The taxonomy is inspired by previous work (Lorenz et al., 2021; Flache et al., 2017). I term the ABMs within social simulation as *deductive ABMs* because they are based on deductions from assumptions of how people change their attitude, as opposed to deduction from empirical observation, which I will discuss in a later section under the name *inductive ABMs*.

**Formal Modeling for Attitude Change in ABMs**

Although a wide variety of formulations of ABMs were proposed in the literature, they can all be viewed as special cases of a single unified formulation that I propose below.

In any given ABM, there are three key functions that fully specify it, the *attitude update function* $(f_{update})$, the *selection function* $(f_{select})$, and the *message function* $(f_{message})$. I will explain the roles of these three functions with the following notations. Assume that there are $N$ agents in a system (each agent has an index $i \in \{1, 2, \ldots, N\} \in \mathbb{N}$; the set is denoted as $I_{system}$), where each agent has an attitude toward the same topic. Let $a_{i,t} \in \mathbb{R}$ be the one-dimensional attitude that an agent $i$ holds at timestep $t \in \{1, 2, \ldots\} = \mathbb{N}$ [1,2]. Let $\mathbb{A}$ be the *attitude space* that $a_{i,t}$ can take value in. When there is no boundary on $a_{i,t}$, the attitude space is the set of real numbers, i.e., $a_{i,t} \in \mathbb{A} = \mathbb{R} = (-\infty, +\infty)$. I will assume $\mathbb{A} = \mathbb{R}$ in the following formulation, but one can also impose boundary on the attitude space, i.e., $\mathbb{A} = [-M, +M]$. Note that the sign of $a_{i,t}$ entails the *direction* of the attitude [3]. That is, $a_{i,t} > 0$ represents a positive attitude (e.g., in favor of a specific policy), and $a_{i,t} < 0$ indicates a negative attitude (e.g., against a specific policy) and $a_{i,t} = 0$ corresponds to a neutral attitude (e.g., neutral to a specific policy). The absolute value of $a_{i,t}$, $|a_{i,t}|$ represents the *magnitude* of the attitude (e.g., how strong the person is in favor of a policy).

---

[1] Not that while most ABMs assume the attitude to be one dimensional, this is not a necessary condition (DeGroot, 1974). Nonetheless, if there are more than one dimensions in the attitude space, one needs to additionally defines what roles different dimensions play in social interaction between agents (e.g., Perfors & Navarro, 2019). This brings an extra layer of complexity, so I defer this to future work and in this present paper, I focus only on the situation where the attitude toward a specific topic takes a single dimension.

[2] Although most ABMs assume the attitude to be a single real number, some models assume attitude to be a distribution over the attitude space, e.g., those who use Bayesian agents to model the change in attitude (Madsen et al., 2018; Perfors & Navarro, 2019). For the sake of simplicity, I will assume attitude to be a single real number but will point out exceptions (where the ABMs assume more than a single real number for the attitude) if that distinction becomes important.

[3] Note that in the literature, some ABMs have an attitude space that is not centered (e.g., $\mathbb{A} = [0,1]$). To ensure that the sign of an attitude entails the direction, one can transform the attitude space so that is centered (e.g., shift $\mathbb{A} = [0,1]$ to $\mathbb{A}' = [-0.5, +0.5]$). In the following formulation, I will assume all attitude space to be centered and symmetric around zero.

Suppose that at timestep $t$, the agent $i$ interacts with $N_J \in \mathbb{N}$ other agents (the set is denoted as $J_{i,t}$) based on a *selection function* ($f_{select}$). Note that $J_{i,t}$ never includes the agent $i$ itself, and the weight that the agent $i$ gives to itself is already encoded in the parameter of *strength of the social influence* (described below). The selected agents then share their messages $M_{i,t} = \{m_{j,t} | j \in J_{i,t}\} \in \mathbb{R}^{N_J}$ with the agent $i$ based on a *message function* ($f_{message}$), which is a function of their attitudes. Formally, $m_{j,t} = f_{message}(a_{j,t})$ for all agent $j$. After the agent $i$ receives the messages $M_{i,t}$, the agent $i$ then updates its attitude from $a_{i,t}$ to $a_{i,t+1}$ based on the *attitude update function* ($f_{update}$). I define $\Delta a_{i,t} = a_{i,t+1} - a_{i,t}$ as the *attitude update* of the agent $i$ from time step $t$ to $t + 1$.

Note that the term that researchers use to refer to *attitude* is not unified in the literature. Common terms include *opinion* (Flache et al., 2017; DeGroot, 1974), *belief* (Madsen et al., 2018), and *attitude* (Lorenz et al., 2021; Hunter et al., 1984). In this review paper, I adopt the term "attitude" because I want to emphasize that it is an internal attitude ($a_{i,t}$) that an agent holds for a specific topic, which is not necessarily the same as the message ($m_{i,t}$) the agent conveys to other agents (as the term "opinion" may imply). Nonetheless, most researchers use these terms interchangeably to refer to the internal attitude. Therefore, what I refer to as "attitude" below should be considered equivalent as "opinion" and "belief".

In the following sections, I will elaborate each of the composite functions of an ABM: attitude update function ($f_{update}$), the selection function ($f_{select}$), and the message function ($f_{message}$).

## Attitude Update Function and Individual Social Influence Function

In this section below, I will define the *attitude update function* ($f_{update}$) and identify its constituent components. Let $f_{update}: A \rightarrow \mathbb{R}$ be the *attitude update function*

$$\Delta a_{i,t} = a_{i,t+1} - a_{i,t} = f_{update}(a_{i,t}, M_{i,t}), \tag{1}$$

where $\Delta a_{i,t}$ is the attitude change of the agent $i$ from time step $t$ to $t+1$, $a_{i,t}$ is the attitude of the agent $i$ *before* receiving the messages, $a_{i,t}$ is the attitude of the agent $i$ *after* receiving the messages, and $M_{i,t}$ are the messages the agent $i$ receive from $J_{i,t}$ (those who interact with the agent $i$ at time step). When there are multiple agents conveying messages to the agent $i$ (i.e., $N_J > 1$), their influence is an aggregation of the social influence of each agent on the agent $i$. I thus rewrite Equation (*1*) as

$$\Delta a_{i,t} = f_{update}(a_{i,t}, M_{i,t}) = \alpha_i \cdot agg\left(\left\{g(a_{i,t}, m_{j,t})\right\}_{j \in J_{i,t}}\right), \qquad (2)$$

where $agg: \mathbb{R}^{N_J} \to \mathbb{R}$ is an aggregation function that summarizes the social influences of all agents in $J_{i,t}$, and $g(a_{i,t}, m_{j,t}): \mathbb{R}^2 \to \mathbb{R}$ is the *individual social influence function* that computes the influence that the message $m_{j,t}$ has on the attitude $a_{i,t}$, and $\alpha_i$ is the *strength of the social influence*, which determines how much the agent gets influenced by the messages it receives. The larger the $\alpha_i$, the more impact other agents have on the agent $i$. Conversely, an agent with higher confidence on its own attitude may have a smaller $\alpha_i$. The term $\alpha_i$ can be a constant, or a function that outputs a real number and the function is independent of $a_{i,t}$ and $M_{i,t}$ (e.g., a function of $N_J$). Note that while most ABMs impose boundary on $\alpha_i$, e.g., $\alpha_i \in [0, 1]$, in theory, $\alpha_i$ can also be negative, where the message pushes the agent's attitude farther way from message.

In the following sections, I will decompose the individual social influence function $g(a_{i,t}, m_{j,t})$ into four components (Flache et al., 2017; Lorenz et al., 2021), where each component is a function of $a_{i,t}$ and/or $m_{i,t}$: 1) assimilation force $asm(a_{i,t}, m_{j,t}): \mathbb{R}^2 \to \mathbb{R}$, 2) reinforcement force $ref(m_{j,t}): \mathbb{R} \to \mathbb{R}$, 3) similarity bias $sim(a_{i,t}, m_{j,t}): \mathbb{R}^2 \to \mathbb{R}_{\geq 0}$, and 4) repulsion force $rep(a_{i,t}, m_{j,t}): \mathbb{R}^2 \to \mathbb{R}$. While these four components were proposed by previous studies (Hunter et al., 1984; Lorenz et al., 2021), to my understanding, this review paper is the first to situate existing ABMs with this framework. In addition, note that most existing ABMs only include a subset of the four components. I can then rewrite the *individual social influence function* as

$$g(a_{i,t}, m_{j,t}) =$$
$$h\Big(asm(a_{i,t}, m_{j,t}), ref(m_{j,t}), sim(a_{i,t}, m_{j,t}), rep(a_{i,t}, m_{j,t})\Big), \tag{3}$$

where $h: \mathbb{R}^4 \to \mathbb{R}$ is a combination function that integrates the four components and outputs a single real number that represents the social influence. For each of the four components, I will elaborate its definition, underlying psychological theories, and one example implementation in the following sections.

### *Assimilation Force*

**Definition**. $asm(a_{i,t}, m_{j,t}): \mathbb{R}^2 \to \mathbb{R}$ is a function that takes the message the agent receives $m_{j,t}$, compare that to the agent's attitude $a_{i,t}$, and outputs the difference (e.g., Euclidean distance). The direction of the assimilation force should be the same as the direction as $(m_{j,t} - a_{i,t})$, i.e., $(m_{j,t} - a_{i,t}) \cdot asm(a_{i,t}, m_{j,t}) > 0$. When there is one single source of influence ($N_{j,t} = 1$), the resultant $\Delta a_{i,t}$ pulls $a_{i,t}$ *towards* $m_{j,t}$ (which is why it is called "assimilation force"), such that $|m_{j,t} - a_{i,t+1}| \le |m_{j,t} - a_{i,t}|$. In addition, the larger the discrepancy between $a_{i,t}$ and $m_{j,t}$ is, the larger the assimilation force should be. That is, the absolute value of $asm(a_{i,t}, m_{j,t})$ is monotonically increasing when $|m_{j,t} - a_{i,t+1}|$ increases, i.e., $|asm(a'_{i,t}, m_{j,t})| \le |asm(a'_{i,t}, m_{j,t})|$ if and only if $|m_{j,t} - a_{i,t+1}| \le |m_{j,t} - a'_{i,t}|$.

**Example.** In DeGroot (1974)'s model (also named "classical averaging model" in the literature; Flache et al., 2017), it assumes the change in attitude is a weighted average of all the differences between the message $m_{j,t}$ and the agent's attitude $a_{i,t}$.

$$\Delta a_{i,t} = \alpha_i \cdot agg\Big(\big\{g(a_{i,t}, m_{j,t})\big\}_{j \in J_{i,t}}\Big) =$$
$$\alpha_i \cdot agg\Big(\big\{asm_{i,t}(a_{i,t}, m_{i,t})\big\}_{j \in J_{i,t}}\Big) = \tag{4}$$

$$\alpha_i \cdot agg(\{(m_{j,t} - a_{i,t})\}_{j \in J_{i,t}}) =$$

$$\alpha_i \cdot \sum_{j \in J_{i,t}} p_{ij}(m_{j,t} - a_{i,t}) =$$

$$\alpha_i \cdot \sum_{j \in J_{i,t}} p_{ij}(a_{j,t} - a_{i,t}),$$

In this model, $g(a_{i,t}, m_{j,t}) = asm(a_{i,t}, m_{i,t}) = (m_{j,t} - a_{i,t})$ and $agg$ is a weighted averaging function, where $p_{ij}$ is the weight of the social influence of the agent $j$ exerts on the agent $i$. Note that in this model, $\sum_{j \in J_{i,t}} p_{ij} = 1$ and $\alpha_i \in [0,1]$. It also assumes that all agents convey their attitudes directly as messages, so $m_{j,t} = f_{message}(a_{j,t}) = a_{j,t}$. Under this attitude update function, it is guaranteed that the updated attitude will be bounded in the range of the agents' attitudes at time $t$, i.e., $\min(a_{i,t+1}, \{a_{j,t} | j \in J_{i,t}\}) \leq a_{i,t+1} \leq \max(a_{i,t+1}, \{a_{j,t} | j \in J_{i,t}\})$.[4]

**Psychological Theories.** Various theories have been cited to account for the assimilation effect, including but not limited to information integration theory (Anderson, 1971), social learning (Akers et al., 1995), cognitive dissonance theory (Festinger, 1962).

*Reinforcement Force*

**Definition.** $ref(m_{i,t})$: $\mathbb{R} \to \mathbb{R}$ is a function that outputs a value that represents the "magnitude" of the message $m_{i,t}$ such that the magnitude is monotonically increasing when $m_{i,t}$ gets larger, under the constraint that it should preserve the sign of $m_{i,t}$. That is, $ref(m_{j,t}) \leq ref(m'_{j,t})$ if and only if $m_{j,t} \leq m'_{j,t}$, and $sign(m_{i,t}) = sign(m_{i,t})$. Some examples include $ref(m_{i,t}) = m_{i,t}$ and $ref(m_{i,t}) = \tanh(m_{i,t})$. Regardless of the attitude of the agent $a_{i,t}$, the message $m_{i,t}$ has an effect to shift $a_{i,t}$ in the same direction as the sign of the message $m_{i,t}$.

---

[1]In DeGroot (1974)'s original formulation, the update rule is specified as $a_{i,t} = \sum_{j \in I_{system}} p'_{i,j} a_{j,t}$, where $\sum_{j \in J_{i,t}} p'_{i,j} = 1$ and $p'_{i,j} \in [0,1]$. Note that it is a weighted average attitude over all agents in the system, including the agent $i$ itself, i.e., $j \in I_{system}$. I rearranged the update rule to $\Delta a_{i,t} = \alpha_i \cdot \sum_{j \in J_{i,t}} p_{ij}(a_{j,t} - a_{i,t})$, where $\sum_{j \in J_{i,t}} p_{ij} = 1$, $p_{ij} \in [0,1]$, $\alpha_i \in [0,1]$, and $p_{ij} = p'_{i,j} \cdot (1/\alpha_i)$ (when $\alpha_i \neq 0$). Note that the formula becomes a weighted average attitude difference over all agents except the agent $i$, i.e., $j \in J_{i,t} = I_{system} \backslash \{i\}$. These two formulas are mathematically equivalent, and the social influence strength $\alpha_i = (1 - p_{ii})$, where $p_{ii}$ can be viewed as "self-weight", the weight the agent $i$ gives to its initial attitude $a_{i,t}$.

**Example.** In Hunter et al. (1984 p. 11, Equation 2.1, and p. 40, Equation 3.4)'s model, it assumes that the attitude change is the sum of the messages the agent receives.

$$\Delta a_{i,t} = \alpha_i \cdot agg\left(\left\{g\left(a_{i,t}, m_{j,t}\right)\right\}_{j \in J_{i,t}}\right) =$$

$$\alpha_i \cdot agg\left(\left\{ref\left(m_{j,t}\right)\right\}_{j \in J_{i,t}}\right) =$$

$$\alpha_i \cdot agg\left(\left\{m_{j,t}\right\}_{j \in J_{i,t}}\right) =$$

$$\alpha_i \cdot \sum_{j \in J_{i,t}} m_{j,t}$$

(5)

In this model, $g\left(a_{i,t}, m_{j,t}\right) = ref\left(m_{j,t}\right) = m_{j,t}$ and $agg$ is a summation function.

**Psychological Theories.** According to the behavioristic learning theories, attitude can be regarded as an agent's evaluative response to a topic. If the message that the agent receives aligns with its response, it reinforces the existing attitude. Other, the message punishes the existing attitude (Doob, 1947; Hunter et al., 1984). Other theories cited for the reinforcement force term includes social contagion theory that considers message as a dose of exposure to a contagious entity like rumors (where they use the term "contagion" to refer to this effect) (Dodds & Watts, 2005; Lorenz et al., 2021).

*Similarity Bias*

**Definition.** $sim\left(a_{i,t}, m_{j,t}\right): \mathbb{R}^2 \to \mathbb{R}_{\geq 0}$ is a non-negative function that monotonically decays as the distance (e.g., Euclidean distance) between the message $m_{j,t}$ and the attitude $a_{i,t}$ increases. That is, $sim\left(a_{i,t}, m_{j,t}\right) \geq sim\left(a_{i,t}, m'_{j,t}\right)$ if and only if $\left|a_{i,t} - m_{j,t}\right| \leq \left|a_{i,t} - m'_{j,t}\right|$.

**Example.** Note that this similarity bias term does not exist alone, and only exists along with other forces (e.g., the assimilation force) to bias its influence. In Hegselmann & Krause (2002)'s bounded confidence (BC) model, they assume that that an agent $j$ has an assimilation force on the agent $i$ only if it is close enough to its attitude $a_{i,t}$.

$$\Delta a_{i,t} = \alpha_i \cdot agg\left(\left\{g\left(a_{i,t}, m_{j,t}\right)\right\}_{j \in J_{i,t}}\right) =$$

$$\alpha_i \cdot agg\left(\left\{h\left(asm\left(a_{i,t}, m_{j,t}\right), sim\left(a_{i,t}, m_{j,t}\right)\right)\right\}_{j \in J_{i,t}}\right) =$$

$$\alpha_i \cdot \sum_{j \in J_{i,t}} sim\left(a_{i,t}, m_{j,t}\right) \cdot asm\left(a_{i,t}, m_{j,t}\right) = \qquad (6)$$

$$\alpha_i \cdot \sum_{j \in J_{i,t}} sim\left(a_{i,t}, m_{j,t}\right) \cdot \left(m_{j,t} - a_{i,t}\right) =$$

$$\frac{N_{\varepsilon_i}}{(N_{\varepsilon_i}+1)} \cdot \frac{1}{(N_{\varepsilon_i})} \cdot \sum_{j \in J_{i,t}} sim\left(a_{i,t}, a_{j,t}\right) \cdot \left(a_{i,t} - a_{j,t}\right),$$

where $\alpha_i = N_{\varepsilon_i}/(N_{\varepsilon_i} + 1)$ for all $i$, the assimilation force $asm\left(a_{i,t}, m_{j,t}\right) = \left(m_{j,t} - a_{i,t}\right)$, $m_{j,t} = f_{message}\left(a_{j,t}\right) = a_{j,t}$ (assuming all agents convey their attitudes as messages), $h$ is a multiplication function, and

$$sim\left(a_{i,t}, a_{j,t}\right) = \begin{cases} 1, \text{if } \left|a_{j,t} - a_{i,t}\right| < \varepsilon_i \\ 0, \text{otherwise} \end{cases}. \qquad (7)$$

In Equation (7), $\varepsilon_i$ is the "confidence bound" of the agent $i$. Whoever falls within this bound has an influence (assimilation force) on the agent $i$. The term $N_{\varepsilon_i}$ in Equation (6) is the number of agents who fall within the confidence bound ($\left|a_{j,t} - a_{i,t}\right| < \varepsilon_i$) of the agent $i$. This makes $agg$ an averaging function over the agents that have influences on the agent $i$. The term $\alpha_i = N_{\varepsilon_i}/(N_{\varepsilon_i} + 1)$ entails that the more agents that fall within the confidence bound, the greater the social influence has on the agent $i$'s attitude change $\Delta a_{i,t}$. Note that this formulation of $\alpha_i$ is not a definitive characteristic of similarity bias, one can also assume that $\alpha_i$ is a constant for a given agent $i$

**Psychological Theories.** Empirical studies have shown that people are more receptive to a given message if it is closer to their original attitude, and different theories have been proposed to explain this phenomena, such as social judgement theory (Sherif & Hovland, 1961), confirmation bias (Nickerson, 1998), and, motivated reasoning (Kunda, 1990). As a result, as the discrepancy between the message and the attitude becomes larger, the influence of the message diminishes.

***Repulsion Force***

**Definition.** Similar to the assimilation function, $rep(a_{i,t}, m_{j,t}): \mathbb{R} \to \mathbb{R}$ is a function that takes the message the agent receives $m_{j,t}$, compare that to the agent's attitude $a_{i,t}$, and computes the difference (e.g., Euclidean distance). Different from the assimilation function, $rep(a_{i,t}, M_{i,t})$ outputs the reverse of this distance. When there is one single source of influence ($N_{j,t} = 1$), the resultant $\Delta a_{i,t}$ pushes $a_{i,t}$ *away* $m_{j,t}$ (which is why it is called "repulsion force"), such that $|m_{j,t} - a_{i,t+1}| \geq |m_{j,t} - a_{i,t}|$. Like the assimilation function, the larger the discrepancy between $a_{i,t}$ and $m_{j,t}$ is, the larger the repulsion force should be. That is, the absolute value of $rep(a_{i,t}, m_{j,t})$ is monotonically increasing when $|m_{j,t} - a_{i,t+1}|$ increases, i.e., $|rep(a_{i,t}, m_{j,t})| \leq |rep(a'_{i,t}, m_{j,t})|$ if and only if $|m_{j,t} - a_{i,t}| \leq |m_{j,t} - a'_{i,t}|$. Conceptually, the repulsion force is the opposite of the assimilation force.

**Example.** Note that the repulsion force does not exist alone, and only exists along with other forces (e.g., the assimilation force). In most ABMs that include the repulsion force, they only specify the case where there is one single source of message at each interaction (i.e., $N_J = 1$). Although there are few exceptions (e.g., Salzarulo, 2006), the formulation is too convoluted to describe succinctly here. Therefore, I will introduce Jager & Amblard (2005)'s social judgement (SJ) model, which applies to the case where $N_J = 1$.

$$\Delta a_{i,t} = \alpha_i \cdot agg\left(\left\{g(a_{i,t}, m_{j,t})\right\}_{j \in J_{i,t}}\right) = \alpha_i \cdot g(a_{i,t}, m_{j,t}) =$$
$$\alpha_i \cdot \left[sim(a_{i,t}, m_{j,t}) \cdot asm(a_{i,t}, m_{j,t}) + rep(a_{i,t}, m_{j,t})\right] = \qquad (8)$$
$$\alpha_i \cdot \left[sim(a_{i,t}, a_{j,t}) \cdot (a_{j,t} - a_{i,t}) + rep(a_{i,t}, a_{j,t})\right],$$

where

$$sim(a_{i,t}, a_{j,t}) \cdot (a_{j,t} - a_{i,t}) = \begin{cases} (a_{j,t} - a_{i,t}), \text{if } |a_{j,t} - a_{i,t}| < u_i \\ 0 \text{ , otherwise} \end{cases} \quad (9)$$

and

$$rep(a_{i,t}, a_{j,t}) = \begin{cases} -(a_{j,t} - a_{i,t}), \text{if } |a_{j,t} - a_{i,t}| > t_i \\ 0 \text{ , otherwise} \end{cases}. \quad (10)$$

In the Equation (*9*) and (*10*), $u_i$ is the threshold for the *latitude of acceptance* and $t_i$ is the threshold for the *latitude of rejection* of the agent $i$, respectively. In this model, agents are assumed to convey their attitudes directly as messages ($m_{j,t} = f_{message}(a_{j,t}) = a_{j,t}$). Note that when $N_J = 1$, this model is equivalent to the bounded confidence model in Equation (*6*) and (*7*) except that Jager & Amblard (2005)'s model assumes the repulsion term as in Equation (*10*).

**Psychological Theories.** The theories that have been cited to justify the repulsive force includes cognitive dissonance theory (Festinger, 1962), social judgment theory (Sherif & Hovland, 1961), self-categorization theory (Salzarulo, 2006). While the inclusion of repulsive force is included according to psychological theories, the empirical evidence with human experiment for this effect is mixed (Takács et al., 2016).

### *Other Components*

While I have summarized the four most common factors that are included in the attitude update functions in the literature, there are some components that I didn't include in my formulation because they are less common. In the section below where I introduce each ABM, if within a model there are components that I didn't include in $g(a_{i,t}, m_{j,t})$ in Equation (*3*), I will highlight them separately. Among those are 1) polarity factor, 2) source credibility, and 3) boundary of attitude, 4) self-decay factor. The polarity factor refers to the tendency that the more extreme one's attitude is (larger $|a_{i,t}|$), the less likely he will change his attitude in social interaction. The polarity factor $pol(a_{i,t})$ has been included in some models (Lorenz et al., 2021; Hunter et al., 1984) as

$$pol\left(a_{i,t}\right) = \frac{M^2 - {a_{i,t}}^2}{M^2} \tag{11}$$

where $M$ is the theoretical boundary for the attitude space (i.e., $a_{i,t} \in \mathbb{A} = [-M, +M]$). Note that if attitude is assumed to be unbounded (e.g., Baumann et al., 2020), $pol\left(a_{i,t}\right)$ needs to be modelled differently. Empirical evidence from the study of congruity theory have been cited to support the role of polarization factor (Lorenz et al., 2021; Osgood & Tannenbaum, 1955). Another component is source credibility $s_{i,j}$, which describes how much the agent $i$ trusts another agent $j$. By definition, $s_{i,j}$ is independent of $a_{i,t}$ and $m_{j,t}$, and is a separate entity that needs to be assumes/measured for each agent. To my understanding, this component is only included in one ABM in the literature (Lorenz et al., 2021). Another factor that varies across ABMs are the boundary of attitude. While most models assume the attitude space to be bounded (e.g., $a_{i,t} \in \mathbb{A} = [-1,1]$), it is not necessary (Baumann et al., 2020). Finally, the self-decay factor is introduced to assume that one's attitude will decay to neutral if there is no social influence, i.e., $a_{i,t} \to 0$ if $t \to 0$. To my knowledge, this self-decay factor is included only in one physics-like ABM in the literature, and is not justified by psychological theories. (Baumann et al., 2020).

**Selection Function and Number of Sources**

Aside from the attitude update function $f_{update}$, another important factor that decides the attitude change $\Delta a_{i,t}$ is the selection function $f_{select}$. The selection function takes as inputs the index of the agent $i$ and the time step $t$, and outputs the set of agents $J_{i,t}$ that will have social influence on the agent $i$. Formally speaking, $f_{select}(i,t): \mathbb{N}^2 \to \mathbb{N}^{N_J}$. The selection function can be driven by internal factors (e.g., people's tendency to interact with those who share similar attitude), or by external factors (e.g., the recommendation algorithms that social media use to deliver messages to users). Some ABMs assume random interactions between agents (e.g., Deffuant et al., 2000), some assumes a underlying graph that defines the influences between agents (e.g., DeGroot, 1974), and some assume that agents tend to interact with those who share similar attitudes (Baumann et al., 2020; Madsen et al., 2018). Different selection functions have been shown to result in distinct macro-level attitude distribution in social simulations even when the agents are using the same attitude update function (Frigo, 2022).

### *Number of Sources*

One subtle yet important factor in the selection function $f_{select}$ is $N_J$, the number of agents in $J_{i,t}$ (the set of agents that have influence on the agent $i$ at $t$). This factor varies across ABMs. For example, some models assume that an agent can only be influenced by one single source at a given time (though the agent can receive influences iteratively from multiple sources at different time steps) (Deffuant et al., 2002, 2000; Madsen et al., 2018), some models that allow influences from multiple sources simultaneously (Hegselmann & Krause, 2002; Salzarulo, 2006), and models require exact two sources at each interaction (Frigo, 2022). The number of sources of influence can be an important factor because it determines the applicability of the ABMs in different interaction settings, and different interaction settings can lead to different macro-level opinion distributions with the same ABM (e.g., sequential update versus parallel update; Salzarulo, 2006; also see Frigo, 2022).

### **Message Function**

The message function determines the message $m_{j,t}$ that an agent $j$ shares based on its attitude $a_{j,t}$. Formally, $m_{j,t} = f_{message}(a_{j,t}): \mathbb{R} \to \mathbb{R}$. Note that most ABMs assume that all agents convey their internal attitude with other agents without bias (i.e., $m_{j,t} = f_{message}(a_{j,t}) = a_{j,t}$), nonetheless, this is not a necessary guarantee (e.g., people may lie deliberately to distort others' attitudes for their own interest). Also note that some models assume that the message $m_{j,t}$ is a random variable based on the agent's attitude distribution, e.g., those who assume agents to be Bayesian learners (Perfors & Navarro, 2019). In their model, the expected value of the $m_{j,t}$ is still equal to the excepted value of the attitude distribution, i.e., $\mathbb{E}(a_{j,t}) = \mathbb{E}(m_{j,t})$.

**Macro-level Attitude Distributions**

The focus of this review paper is to compare different families of ABMs, rather than bridging the micro-level ABMs with macro-level phenomenon (for a review, see Flache et al., 2017; Lorenz et al., 2021). Nonetheless, the relevance to different macro-level attitude distributions is an important characteristic of an ABM. The reason is that the deductive ABMs are proposed with the intent to explain the macro-level attitude distribution (e.g., bipolarization, echo chamber, extremism, consensus formation). In the section below, I will briefly review the common attitude distributions that the ABMs attempt to explain. Attitude distribution, $f_{A,t}(a)$, refers to the probability mass function (over the attitude space $\mathbb{A}$) of the attitudes of all the $N$ agents in the system ($I_{system}$) at time step $t$. Different taxonomies and terminologies have been used to classify attitude distributions. I adopt the taxonomy suggested in Flache et al. (2017) because it is simple and comprehensive. I refer readers who are interested in a more nuanced taxonomy of different attitude distributions to Lorenz et al (2021).

*Consensus*

Consensus, also referred to as *uniformity* (Jager & Amblard, 2005)*,* is characterized by a unimodal distribution where the attitudes of the agents tend to agree with each other.

*Extremization*

Extremization, also referred to as radicalization (Baumann et al., 2020), extreme consensus (Lorenz et al., 2021), group polarization, is a special case of consensus where the mode of the consensus is far away from the neutral attitude. Formally, it can be defined by $|mode(f_{A,t}(a))| \geq \varepsilon$, where $\varepsilon$ is the threshold of extremization. While most studies do not explicitly specify the value of $\varepsilon$ (Lorenz et al., 2021 is an exception), when the attitude space is bounded, $f_{A,t}(a)$ is often regarded as extremization if the mode is close to the attitude boundary ($-M$ or $+M$).

### Fragmentation

Fragmentation, also referred to as pluriformity (Jager & Amblard, 2005), clustering (Flache et al., 2017), echo chambers (Baumann et al., 2020; Madsen et al., 2018; Perfors & Navarro, 2019), is characterized by at least two modes in the distribution $f_{A,t}(a)$.

### Bipolarization

Bipolarization is a special case of fragmentation where there is exact two modes in the distribution $f_{A,t}(a)$, and the two nodes are of different signs (i.e., one mode is positive, and another mode is negative), i.e., $mode_1(f_{A,t}(a)) \cdot mode_2(f_{A,t}(a)) < 0$. In addition, each of the two modes should be far away from the neutral attitude. While few studies formally define how far way each mode should be from the neutral attitude (Lorenz et al., 2021 is an exception), when the attitude space is bounded, $f_{A,t}(a)$ is often regarded as bipolarization if the two modes are close to the different sides of attitude boundary ($-M$ and $+M$).

## Empirical Evidence of Deductive ABMs

Because the goal of deductive ABMs is to explain the formation of different attitude distribution among people, any ABM should be validated against empirical data. There are two primary ways that can validate an ABM (Flache et al., 2017). First, one can conduct human experiments where subjects interact with another and measure how their attitudes change throughout the interaction. The empirical results can then provide insights for the formulation of ABMs. I term this this type of ABMs as *inductive ABMs* and will review them in a separate section below. A second way to validate an inductive ABM is to compare its macro-level simulated attitude distributions against empirical attitude distributions (e.g., survey data). If a deductive ABM fails to simulate some common empirical attitude distributions, then the ABM would be less supportive compared to other ABMs that are capable of. On the other hand, if an ABM can simulate empirical attitude distributions, then it suffices the *necessary* condition for the model to be valid. Note that this type of evidence is not *sufficient* because the exact same macro-level attitude distribution can be simulated by different underlying ABMs. Surprisingly, most deductive ABMs are not validated against any empirical data, although most of them claim to model what a human is supposed to do.

**The Gallery of the Deductive ABMs**

In the following section, I will enumerate the deductive ABMs and describe their attitude update functions, selection functions, the macro-level attitude distributions simulated based on the ABMs, and their empirical evidence (also summarized in Table 1.).

Table 1.

*Inductive ABMs that model attitude update.*

| | Update Function | | | | | # sources $(N_J)$ | Selection Function $(f_{select})$ | Macro-level Attitude Distribution |
|---|---|---|---|---|---|---|---|---|
| Study | asm. | ref. | sim. | rep. | Other | | | |
| (DeGroot, 1974) *Classic Averaging Model* | O | X | X | X | X | $1 \leq N_J \leq N$ | All connected neighbors | Consensus |
| (Deffuant et al., 2000; Flache et al., 2017) *Bounded Confidence (BC) Model* | O | X | O | X | X | $N_J = 1$ | Any random agent within the bound | Consensus Fragmentation Bipolarization Extremism |

| | | | | | | $N_J$ | | |
|---|---|---|---|---|---|---|---|---|
| (Hegselmann & Krause, 2002) *Bounded Confidence (BC) Model* | O | X | O | X | X | $1 \leq N_J \leq N$ | All agents within the bound | Consensus Fragmentation |
| (Deffuant et al., 2002) *Relative Agreement (RA) Model* | O | X | O | X | X | $N_J = 1$ | Any random agent | Consensus Bipolarization Extremism |
| (Jager & Amblard, 2005) *Social Judgement (SJ) Model* | O | X | X | O | X | $N_J = 1$ | Any random agent | Consensus Fragmentation Bipolarization |
| (Salzarulo, 2006) | O | X | X | O | X | $1 \leq N_J \leq N$ | All neighbors that are "visible" to the agent | Consensus Fragmentation Bipolarization |
| (Lorenz et al., 2021) | O | O | O | X | pol. src. | $N_J = 1$ | Any random agent | Consensus Fragmentation Bipolarization Extremism |
| (Madsen et al., 2018) | O | X | O | X | X | $N_J = 1$ | Any random agent within the bound | Fragmentation |
| (Baumann et al., 2020) | X | O | O built in $f_{select}$ | X | sdc. | $1 \leq N_J \leq N$ | Any agent $j$ that agent $i$ contacts according to "homophily" | Consensus Bipolarization Extremism |

*Note:* asm. = assimilation force; sim. = similarity bias; ref. = reinforcement force; rep. = repulsion force; pol. = polarization factor; src. = source credibility; sdc. = self-decay factor

*Opinion Dynamics: DeGroot (1974)'s Model*

DeGroot (1974) proposed a classic averaging model which assumes that the change in attitude is a weighted average of all the differences between the message $m_{j,t}$ and the agent's attitude $a_{i,t}$.

**Attitude update function.** The attitude update function is as follows.

$$\Delta a_{i,t} = \alpha_i \sum_{j \in J_{i,t}} p_{ij} (a_{j,t} - a_{i,t}), \tag{12}$$

- **Aggregation Function.** A weighted average function. Formally, $agg\left(\left\{g\left(a_{i,t}, m_{j,t}\right)\right\}_{j \in J_{i,t}}\right) = \sum_{j \in J_{i,t}} p_{ij} g\left(a_{i,t}, m_{j,t}\right)$, where $p_{ij} \in [0,1]$ and $\sum_{j \in J_{i,t}} p_{ij} = 1$.

- **Assimilation Force.** $asm\left(a_{i,t}, m_{i,t}\right) = \left(m_{j,t} - a_{i,t}\right)$.

- Reinforcement Force. None.

- Similarity Bias. None.

- Repulsion Force. None.

- Other Assumptions. (1) $\alpha_i \in [0,1]$.

**Selection Function.** $J_{i,t} = f_{select}(i, t) = \{$the agents $j$ whose $p_{ij} > 0\}$

- Number of Sources. $1 \leq N_J \leq N$.

**Message Function.** $m_{j,t} = f_{message}\left(a_{j,t}\right) = a_{j,t}$

**Macro-level Attitude Distribution.** Consensus or segmentation, depending on the distribution of $p_{ij}$.

Empirical Evidence. None.

*Opinion Dynamics: Deffuant et al. (2000)'s Model*

Deffuant et al. (2000) proposed a bounded-confidence (BC) model for the case of $N_J = 1$. It assumes that if the message $m_{j,t}$ is close enough to the agent $i$'s attitude $a_{i,t}$, the message has an assimilation force on the agent's attitude.

**Attitude update function.** The attitude update function is as follows.

$$\Delta a_{i,t} = \alpha_i \cdot sim(a_{i,t}, m_{j,t}) \cdot (a_{j,t} - a_{i,t}), \tag{13}$$

where

$$sim(a_{i,t}, m_{j,t}) = \begin{cases} 1, \text{if } |a_{j,t} - a_{i,t}| < \varepsilon_i \\ 0, \text{otherwise} \end{cases}. \tag{14}$$

- *Aggregation Function.* No aggregation function because $N_J = 1$.

- **Assimilation Force.** $asm(a_{i,t}, m_{j,t}) = (m_{j,t} - a_{i,t})$.

- Reinforcement Force. None.

- **Similarity Bias.** $sim(a_{i,t}, m_{j,t}) = \begin{cases} 1, \text{if } |a_{j,t} - a_{i,t}| < \varepsilon_i \\ 0, \text{otherwise} \end{cases}$, where $\varepsilon_i$ is the "confidence bound" of the agent $i$.

- Repulsion Force. None.

- Other Assumptions. (1) $\alpha_i \in [0,1]$.

**Selection Function.** $J_{i,t} = f_{select}(i, t) = \{$one random agent $j$ in the system within the confidence bound, i.e., whose $|a_{j,t} - a_{i,t}| < \varepsilon_i$, except the agent $i$ itself$\}$.

- Number of Sources. $N_J = 1$.

**Message Function.** $m_{j,t} = f_{message}(a_{j,t}) = a_{j,t}$

**Macro-level Attitude Distribution.** Consensus and fragmentation. If there are special "extremist" (agents with smaller $\varepsilon$ ) and extremists in initial attitude distribution ($f_A(a)$ at $t = 1$), bipolarization, and extremization can occur (Flache et al., 2017).

Empirical Evidence. None.

*Opinion Dynamics: Hegselmann & Krause (2002)'s Model*

Hegselmann & Krause (2002)'s bounded confidence (BC) model can be regarded as an extension of Deffuant et al. (2000)'s model in that it can handle multiple sources, i.e., $1 \leq N_J \leq N$.

**Attitude update function.** The attitude update function is as follows.

$$\Delta a_{i,t} = \frac{N_{\varepsilon_i}}{(N_{\varepsilon_i}+1)} \cdot \frac{1}{(N_{\varepsilon_i})} \cdot \sum_{j \in J_{i,t}} sim(a_{i,t}, a_{j,t}) \cdot (a_{i,t} - a_{j,t}), \tag{15}$$

where

$$sim(a_{i,t}, m_{j,t}) = \begin{cases} 1, \text{if } |a_{j,t} - a_{i,t}| < \varepsilon_i \\ 0, \text{otherwise} \end{cases}. \tag{16}$$

- **Aggregation Function.** An average function over all who fall within the confidence bound of the agent $i$. Formally, $agg\left(\{g(a_{i,t}, m_{j,t})\}_{j \in J_{i,t}}\right) = \frac{1}{(N_{\varepsilon_i})} \cdot \sum_{j \in J_{i,t}} g(a_{i,t}, m_{j,t})$, where $N_{\varepsilon_i}$ are the number of agents within the confidence bound.

- **Assimilation Force.** $asm(a_{i,t}, m_{j,t}) = (m_{j,t} - a_{i,t})$.

- Reinforcement Force. None.

- **Similarity Bias.** $sim(a_{i,t}, m_{j,t}) = \begin{cases} 1, \text{if } |a_{j,t} - a_{i,t}| < \varepsilon_i \\ 0, \text{otherwise} \end{cases}$, where $\varepsilon_i$ is the confidence bound of the agent $i$.

- Repulsion Force. None.

- *Other Assumptions.* The strength of the social influence is a function of the number of agents in the confidence bound. Formally, $\alpha_i(N_{\varepsilon_i}) = \frac{N_{\varepsilon_i}}{(N_{\varepsilon_i}+1)}$. Note that $\alpha_i(N_{\varepsilon_i})$ is monotonically increasing when $N_{\varepsilon_i}$ gets larger and it is bounded within $[0.5, 1]$. This assumption can be interpreted as that the agent $i$ give an equal weight to all the agents within the confidence bound including the agent $i$ itself (i.e., equal weight to all agents in the set $\{1 \leq j \leq N \mid |a_{j,t} - a_{i,t}| < \varepsilon_i\}$).

Therefore, the larger $N_{\varepsilon_i}$ is, the less the agent $i$ gives weight to itself, and thus $\alpha_i(N_{\varepsilon_i})$ approaches 1.

**Selection Function.** $J_{i,t} = f_{select}(i,t) = I_{system} \backslash \{i\}$ (all the $N$ agents in the system except the agent $i$). Note that only those within the confidence bound get to influence the agent $i$'s attitude.

- Number of Sources. $N_J = N - 1$.

**Message Function.** $m_{j,t} = f_{message}(a_{j,t}) = a_{j,t}$

Macro-level Attitude Distribution. Consensus, fragmentation.

Empirical Evidence. None.

### *Opinion Dynamics: Deffuant et al. (2002)'s Model*

Deffuant et al. (2002) proposed the relative agreement (RA) model which extends the bounded-confidence model in that the similarity bias $\mathrm{a}sm(a_{i,t}, m_{j,t})$ is a continuously decaying function rather than a step function (as in the BC model, Equation (*14*)). Note that this model assumes $N_J = 1$. The rationale under this model is that the more the message $m_{j,t}$ agrees with $a_{i,t}$, the more susceptible the agent's $i$ is to its assimilation force.

Note that in the RA model, it assumes that there is uncertainty around each agent's attitude $a_{i,t}$. The uncertainty is modeled by the term $u_{i,t} > 0$. Therefore, the agent $i$'s attitude is modeled as a segment $seg_{i,t} = [a_{i,t} - u_{i,t}, a_{i,t} + u_{i,t}]$, and the message $m_{j,t}$ conveyed by the agent $j$ is another segment $seg_{j,t} = [a_{j,t} - u_{j,t}, a_{j,t} + u_{j,t}]$. When the agent $j$ attempts to influence the agent $i$, the larger these two segments "agree" (see below for how this is quantified) , the larger the influence should be. Because the attitude now comes with the uncertainty term, the similarity bias function $sim(a_{i,t}, m_{j,t})$ should be modified to $sim(a_{i,t}, u_{i,t}, a_{j,t}, u_{j,t}): \mathbb{R}^4 \rightarrow \mathbb{R}_{\geq 0}$.

**Attitude update function.** The attitude update function is as follows.

$$\Delta a_{i,t} = \alpha_i \cdot sim(a_{i,t}, u_{i,t}, a_{j,t}, u_{j,t}) \cdot (a_{i,t} - a_{j,t}), \tag{17}$$

where

$$sim(a_{i,t}, u_{i,t}, a_{j,t}, u_{j,t}) = \begin{cases} (h_{i,j}/u_j) - 1, \text{if } (h_{i,j}/u_j) > 1 \\ 0, \text{otherwise} \end{cases}. \tag{18}$$

In Equation (*18*), $h_{i,j}$ is the overlap between $seg_{i,t}$ and $seg_{j,t}$, i.e., $h_{i,j} = \min(a_{i,t} + u_{i,t}, a_{j,t} + u_{j,t}) - \min(a_{i,t} - u_{i,t}, a_{j,t} - u_{j,t})$. The term $(h_{i,j}/u_j) - 1$ is referred to as the relative agreement of the agent $j$ with the agent $i$. Note that the uncertainty term $u_{i,t}$ is assumed to update in a similar way as in Equation (*17*), (*18*) (for details, see Deffuant et al., 2002).

- *Aggregation Function.* No aggregation function because $N_J = 1$.

- **Assimilation Force.** $asm(a_{i,t}, m_{j,t}) = (m_{j,t} - a_{i,t})$.

- Reinforcement Force. None.

- **Similarity Bias.** $sim(a_{i,t}, u_{i,t}, a_{j,t}, u_{j,t}) = \begin{cases} (h_{i,j}/u_j) - 1, \text{if } (h_{i,j}/u_j) > 1 \\ 0, \text{otherwise} \end{cases}$,

  where $h_{i,j}$ is the overlap between $seg_{i,t}$ and $seg_{j,t}$.

- Repulsion Force. None.

- *Other Assumptions.* (1) $\alpha_i \in [0,1]$. (2) As described above, the agent $i$'s attitude is modeled as a segment $seg_{i,t} = [a_{i,t} - u_{i,t}, a_{i,t} + u_{i,t}]$, and the message $m_{j,t}$ conveyed by the agent $j$ is another segment $seg_{j,t} = [a_{j,t} - u_{j,t}, a_{j,t} + u_{j,t}]$

**Selection Function.** $J_{i,t} = f_{select}(i, t) = \{$one random agent $j$ in $I_{system}$, except the agent $i$ itself $\}$.

- Number of Sources. $N_J = 1$.

**Message Function.** $m_{j,t} = f_{message}([a_{j,t} - u_{j,t}, a_{j,t} + u_{j,t}]) = [a_{j,t} - u_{j,t}, a_{j,t} + u_{j,t}]$

**Macro-level Attitude Distribution.** Consensus, bipolarization, and extremization. Note that compared to the BC model, the RA model can simulate extremization and bipolarization more easily (Deffuant et al., 2002).

Empirical Evidence. None.

### Opinion Dynamics: Jager & Amblard (2005)'s Model

Jager & Amblard (2005) proposed the social judgement (SJ) model, which is similar to the BC model in that there is also a confidence bound (they termed it the *latitude of acceptance*) and the assimilation force, but differs from the BC model in that the SJ model additionally includes a repulsion force. Conceptually, the SJ model assumes that an agent will *assimilate towards* the message $m_{j,t}$ if it is close enough to its attitude $a_{i,t}$ (within the *latitude of acceptance*, as in the BC model), and will *distance away* from the message $m_{j,t}$ if it is too far away from its attitude $a_{i,t}$ (within the *latitude of rejection*). Note that this model assumes $N_J = 1$.

**Attitude update function.** The attitude update function is as follows.

$$\Delta a_{i,t} = \alpha_i \cdot \left[ sim(a_{i,t}, m_{j,t}) \cdot (a_{j,t} - a_{i,t}) + rep(a_{i,t}, m_{j,t}) \right], \tag{19}$$

where

$$sim(a_{i,t}, a_{j,t}) \cdot (a_{j,t} - a_{i,t}) = \begin{cases} (a_{j,t} - a_{i,t}), \text{if } |a_{j,t} - a_{i,t}| < u_i \\ 0, \text{otherwise} \end{cases} \tag{20}$$

and

$$rep(a_{i,t}, a_{j,t}) = \begin{cases} -(a_{j,t} - a_{i,t}), \text{if } |a_{j,t} - a_{i,t}| > t_i \\ 0, \text{otherwise} \end{cases}. \tag{21}$$

In the Equation ($20$) and ($21$), $u_i$ is the threshold for the latitude of acceptance, and $u_i$ is the thresholds for the latitude of rejection of the agent $i$, respectively.

- ***Aggregation Function.*** No aggregation function because $N_J = 1$.

- **Assimilation Force.** $asm(a_{i,t}, m_{j,t}) = (m_{j,t} - a_{i,t})$.

- Reinforcement Force. None.

- **Similarity Bias.** $sim(a_{i,t}, a_{j,t}) = \begin{cases} 1, \text{if } |a_{j,t} - a_{i,t}| < u_i \\ 0, \text{otherwise} \end{cases}$, where $u_i$ is the threshold for the latitude of acceptance of the agent $i$.

- **Repulsion Force.** $rep(a_{i,t}, a_{j,t}) = \begin{cases} -(a_{j,t} - a_{i,t}), \text{if } |a_{j,t} - a_{i,t}| > t_i \\ 0, \text{otherwise} \end{cases}$, where $t_i$ is the threshold for the latitude of rejection of the agent $i$.

- *Other Assumptions.* (1) $\alpha_i \in [0,1]$. (2) As described above, each agent $i$ has two thresholds, $u_i$ and $t_i$, that specify the latitude of acceptance and the latitude of rejection, respectively.

**Selection Function.** $J_{i,t} = f_{select}(i,t) = \{$one random agent $j$ in $I_{system}$, except the agent $i$ itself$\}$.

- Number of Sources. $N_J = 1$.

**Message Function.** $m_{j,t} = f_{message}(a_{j,t}) = a_{j,t}$

**Macro-level Attitude Distribution.** Consensus, fragmentation, bipolarization. Note that compared to the BC model and the RA model, SJ model can simulate extremization and bipolarization without assuming excessive agents at the extremes in the initial attitude distribution at $t = 1$, or special extremist agents that have smaller confidence bound than the other.

Empirical Evidence. None.

***Opinion Dynamics: Lorenz et al. (2021)'s Model***

Lorenz et al. (2021) proposed a model which includes assimilation force, reinforcement force, similarity bias, polarization factor, and source credibility.

**Attitude update function.** The attitude update function is as follows.

$$\Delta a_{i,t} = s(i,j) \frac{\lambda^k}{\lambda^k + |m_{j,t} - a_{i,t}|^k} \frac{(M^2 - a_{i,t}^2)}{M^2} \alpha_i (m_{j,t} - \rho a_{i,t}) = \tag{22}$$

$$\alpha_i s(i,j) pol(a_{i,t}) sim(a_{i,t}, m_{j,t}) \cdot [\rho \cdot asm(a_{i,t}, m_{j,t}) + (1-\rho) \cdot$$

$$ref(m_{j,t})],$$

where each component is defined as follows.

$$pol(a_{i,t}) = \frac{M^2 - a_{i,t}^2}{M^2} \tag{23}$$

The hyperparameter parameter $M$ is the theoretical boundary for the attitude space (i.e., $a_{i,t} \in \mathbb{A} = [-M, +M]$).

$$sim(a_{i,t}, m_{j,t}) = \frac{\lambda^k}{\lambda^k + |m_{j,t} - a_{i,t}|^k} \tag{24}$$

The hyperparameters $\lambda$ and $k$ specify the shape of the similarity bias function.

$$ref(m_{j,t}) = m_{j,t} \tag{25}$$

$$asm(a_{i,t}, m_{j,t}) = (m_{j,t} - a_{i,t}) \tag{26}$$

In Equation (22), $\rho$ is the *degree of assimilation*, which controls the relative contribution of the assimilation force versus the reinforcement force. When $\rho = 1$ there is only assimilation force, and when $\rho = 0$ there is only reinforcement force.

- ***Aggregation Function.*** No aggregation function because $N_J = 1$.

- **Assimilation Force.** $asm(a_{i,t}, m_{j,t}) = (m_{j,t} - a_{i,t})$

- **Reinforcement Force.** $ref(m_{j,t}) = m_{j,t}$

- **Similarity Bias.** $sim(a_{i,t}, m_{j,t}) = \frac{\lambda^k}{\lambda^k + |m_{j,t} - a_{i,t}|^k} \in [0,1]$

- Repulsion Force. None.

- **Other Assumptions.** (1) $\alpha_i \in [0,1]$. Moreover, they also assume $\alpha_i = \alpha$ for all $i$. (2) source credibility $s(i,j) \in [0,1]$, (3) polarization factor $pol(a_{i,t}) = \frac{M^2 - a_{i,t}^2}{M^2} \in [0,1]$

**Selection Function.** $J_{i,t} = f_{select}(i,t) = \{$one random agent $j$ in $I_{system}$, except the agent $i$ itself$\}$.

- Number of Sources. $N_J = 1$.

**Message Function.** $m_{j,t} = f_{message}(a_{j,t}) = a_{j,t}$

**Macro-level Attitude Distribution.** Depending on the values of the hyperparameters $\lambda$, $k$, and $\rho$, all the following distributions are possible: consensus, fragmentation, bipolarization, extremization.

**Empirical Evidence.** They compared simulated attitude distributions against empirical attitude distributions in social survey data. The comparison showed that the simulations based on their ABMs can more realistically resemble the empirical distributions than some other ABMs (e.g., the BC model, the SJ model).

In the section below, I will introduce some ABMs that are not easily expressed by Equation (*2*) and (*3*). Nonetheless, one can still identify the some of the four common components in their formulation. This includes 1) Bayesian agents, and 2) physics-like models.

### Bayesian Agents: Madsen et al (2018)'s Model

Madsen et al (2018) proposed an Bayesian learning agents that updates its attitude $a_{i,t}$ after receiving a message $m_{j,t}$ from another agent $j$. In this model, it assumes that any agent $i$ maintains a normal distribution over the attitude space, i.e., $P(a_{i,t}) = N(\mu_{i,t}, \sigma_{i,t})$. When another agent $j$ interacts with $i$, it conveys its mean $\mu_{j,t}$ as the message, i.e., $m_{j,t} = f_{message}(P(a_{i,t})) = \mu_{j,t}$.

**Attitude update function.** After the agent $i$ receives the message $m_{j,t} = \mu_{j,t}$. , it updates its attitude using the Bayesian rule, i.e.,

$$P(a_{i,t+1}) = P(a_{i,t}|m_{j,t}) = \qquad (27)$$

$$P\left(a_{i,t}\right) \cdot \frac{P\left(m_{j,t}\middle|N\left(\mu_{i,t}, \sigma_{i,t}\right)\right)}{P(E)} = P\left(a_{i,t}\right) \cdot \frac{P\left(\mu_{j,t}\middle|N\left(\mu_{i,t}, \sigma_{i,t}\right)\right)}{P(E)},$$

where $P\left(a_{i,t+1}\right)$ is the attitude distribution at time step $t+1$, which is the same as $P\left(a_{i,t}|m_{j,t}\right)$, the posterior attitude distribution after receiving $m_{j,t}$. $P\left(m_{j,t}\middle|N\left(\mu_{i,t}, \sigma_{i,t}\right)\right)$ is the likelihood function of observing $m_{j,t}$ given the agent $i$'s current attitude distribution, and $P(E)$ is normalizer, which is the marginal probability that $m_{j,t}$ occurs irrespective to the any attitude distribution. In addition, the ABM also assumes a "pruning parameter" $\beta$ such that an agent $i$ gets influenced by another agent $j$ only if their attitudes are not too different (like the idea of confidence bound in the BC model). Formally, $j$ should satisfy $(\mu_i - \beta\sigma_i) \leq \mu_j \leq (\mu_i + \beta\sigma_i)$ to change the agent $i$'s attitude $P\left(a_{i,t}\right)$.

- *Aggregation Function.* No aggregation function because $N_J = 1$.

- **Assimilation Force.** $asm\left(\mu_{i,t}, m_{j,t}\right)$ should be in the same direction as $\left(m_{j,t} - \mu_{i,t}\right)$. That is, $asm\left(\mu_{i,t}, m_{j,t}\right) \cdot \left(m_{j,t} - \mu_{i,t}\right) > 0$.

- Reinforcement Force. None.

- *Similarity Bias.* $j$ should satisfy $(\mu_i - \beta\sigma_i) \leq \mu_j \leq (\mu_i + \beta\sigma_i)$ to have an influence on the agent $i$'s attitude $P\left(a_{i,t}\right)$.

- Repulsion Force. None.

- Other Assumptions. (1) $\alpha_i \in \mathbb{A} = \mathbb{R}$.

**Selection Function.** $J_{i,t} = f_{select}(i, t) = \{$one random agent $j$ from those that is connected with the agent $i$ in a social network$\}$. The social network is either a randomly connected network, or a scale-free network.

- Number of Sources. $N_J = 1$.

**Message Function.** $m_{j,t} = f_{message}\left(P\left(a_{j,t}\right)\right) = \mu_j$

Macro-level Attitude Distribution. Fragmentation.

Empirical Evidence. None.

***Physics-like model: Baumann et al. (2020)'s Model***

Baumann et al. (2020) proposed a model that defines attitude change of all agents in the system with $N$ coupled ordinary differential equations. For any agent $i$, its attitude change is modelled as follows.

**Attitude update function.** The differential equation that defines attitude update is as follows.

$$\frac{da_{i,t}}{dt} = -a_i + \alpha_i \sum_{j \in J_{i,t}} \tanh(c \cdot a_{j,t}), \qquad (28)$$

where $-a_i$ is the *self-decaying factor*, $\tanh: \mathbb{R} \to [-1, +1]$ is the hyperbolic tangent function, $c > 0$ is the *controverlsialiness factor* that controls the shape of the the tanh function. Note that the tanh function is assumed so that the individual social influence is capped within $[-1, +1]$. The greate $c$ is, the steeper the function is. That is, even agent with moderate attitude can have a strong influence on the agent $i$. The self-decaying funcion is included with the assumption that an agent's attitude should decay to neutral without social influences, i.e., $a_{i,t} \to 0$ if $t \to 0$.

- **Aggregation Function.** A summation function. Formally,
$$agg\left(\left\{g\left(a_{i,t}, m_{j,t}\right)\right\}_{j \in J_{i,t}}\right) = \sum_{j \in J_{i,t}} g\left(a_{i,t}, m_{j,t}\right).$$

- Assimilation Force. None.

- **Reinforcement Force.** $ref\left(m_{j,t}\right) = \tanh(c \cdot m_{j,t})$. $c > 0$ is the controverlsialiness factor.

- Similarity Bias. None.

- Repulsion Force. None.

- Other Assumptions. (1) $\alpha_i \in \mathbb{A} = \mathbb{R}$.

**Selection Function.**  At each time step, an agent is activated with probability $act_i$ (sampled from a distribution $F(act) \sim act^{-\gamma}$ that follows a power law). If the agent $i$ is activated, it will contact another the agent $j$ ($i \neq j$) with probability $p_{i,j} = \frac{|a_i - a_j|^{-\beta}}{\sum_{j \in I_{system}} |a_i - a_j|^{-\beta}}$, where $J_{i,t} = f_{select}(i, t) = \{$all agents that are contacted by $i$ at time step $t \}$. The hyperparameter $\beta$ controls the *homophilic effect* in the selection function. The larger the value of $\beta$, the more likely an agent will select a likeminded agent to interact with.

- Number of Sources. $1 \leq N_J \leq N$.

**Message Function.** $m_{j,t} = f_{message}(a_{j,t}) = a_{j,t}$

Macro-level Attitude Distribution.  Fragmentation, Bipolarization.

**Empirical Evidence.** They compared simulated attitude distributions against empirical attitude distributions on three datasets of polarized debates on Twitter. They showed that the simulations based on their ABM resemble the attitude distributions on the Twitter datasets. Nonetheless, they did not compare the simulations of their ABM with simulations based on other ABMs. Therefore, it is hard to tell if their ABM is better at simulating empirical attitude distributions than the other.

**Discussion on Deductive ABMs**

In the section above, I propose a general formulation of ABM, which includes an attitude update function, a selection function, and a message function. In addition, I identify four common factors in the attitude update function. With the general formulation, I then review existing deductive ABMs from the literature of opinion dynamics, Bayesian agents, and physics-like models. To my knowledge, this review is the first to show that all existing ABMs can be viewed as special cases of the general formulation.

There are major limitations with deductive ABMs, and these constraints prevent these ABMs from producing useful predictions. First, most ABMs are not verified against empirical data. Even when macro-level verification is done by comparing simulations with empirical attitude distribution, the comparison is not done in a systematic way, e.g., only include a few ABMs to compare with (Baumann et al., 2020; Lorenz et al., 2021). Second, while most deductive ABMs focus on the formulation of the attitude update function, the choices of selection function, message function, and aggregation function have rarely been justified. For instance, some ABMs assume that an agent can only interact only with one agent at a time. It is thus unknown what should the agent update its attitude when interacting with multiple agents at the same time.

While there is a lack of empirical justifications for most deductive ABMs, there are human behavioral studies that provide evidence for how humans update their attitude through social interaction. While these experiments tend to be small-scale (e.g., less than 50 people in a system), or involves "artificial agents" interacting with the subject, they provide insights for the formulation of ABMs. I term this this type of *inductive ABMs* and will review them below.

## Inductive ABMs: Human Experiments

Apart from the deductive ABMs, a separate paradigm of using ABMs in attitude change is to induce attitude update rules from human experiments (hence the name *inductive ABMs*). The goal of using inductive ABMs is to understand how humans *actually* update their attitudes under social influences. Under this category, there are two primary approaches of applying an ABM, 1) hypothesis testing, and 2) data-driven modeling. I will delineate these two approaches below with concrete examples.

### Hypothesis Testing: Wisdom of Crowds Effect

One reason of using ABMs is to generate predictions based on hypotheses and design human experiments to support or falsify the hypotheses. For example, Becker et al. (2019) proposed an ABM based on the *group polarization* hypothesis and showed that, contrary to the model prediction, people reduced their polarization after interacting with like-minded individuals.

A specific topic that has benefited from ABMs in hypothesis testing is the wisdom of crowds (WOC) effect under social influence: if there is a ground-truth attitude, do people improve their accuracies after exposing to others' messages? For example, suppose each individual in a group is asked about the current population in the world (the ground truth, $a^*$, is around 8 billion). After each person gives their initial estimate ($a_{i,t}$ for all $i$), they are then asked to exchange their messages to one another ($m_{i,t}$ for all $i$). Then, everyone is asked again about the same question ($a_{i,t+1}$ for all $i$). The *wisdom of crowd* (WOC) effect describes the observations that at any given time step $t$, the median of the attitude distribution median $\left(f_{A,t}(a)\right)$ is closer to $a^*$ than most individual's attitude $a_i$ (Galton, 1907).

Studies on the effect of social influences on WOC found that after exchange of messages between people, 1) the median of the updated attitude distribution becomes closer to $a^*$, i.e., $\left|\text{median}\left(f_{A,t+1}(a)\right) - a^*\right| < \left|\text{median}\left(f_{A,t}(a)\right) - a^*\right|$, and 2) the variance of the updated attitude distribution becomes smaller, i.e., $\text{variance}\left(f_{A,t+1}(a)\right) < \text{variance}\left(f_{A,t}(a)\right)$ (Becker et al., 2017, 2019; Jayles et al., 2017). I will elaborate the studies in this literature below, including their experimental designs, the ABM formulations, and empirical findings.

### Becker et al. (2017)

They hypothesized that social influence could improve the WOC effect because the more accurate a person's initial attitude was (i.e., smaller $|a_{i,t} - a^*|$), the more confident he was in this own initial attitude, and thus was less susceptible to the social influences (i.e., having a smaller $\alpha_i$ value), whereas those whose attitudes are more erroneous would be more receptive to social influence. Collectively, this would improve the WOC effect after people exert social influence on one another, i.e., $\left|\text{median}\left(f_{A,t+1}(a)\right) - a^*\right| < \left|\text{median}\left(f_{A,t}(a)\right) - a^*\right|$.

**Experimental Design.** They recruited 1360 subjects and randomly assigned them into 34 separate groups. Each group had 40 subjects who interacted with each other. Each group belonged to one of the three conditions, the decentralized network condition (13 groups), the centralized network condition (13 groups), and the control condition (8 groups). Across all three conditions, all subjects went through three rounds of estimation tasks ($t = 1, 2, 3$) for several different topics. In each estimation task, each subject provides an estimate about a question (e.g., "How many candies are there in the container" along with an image), denoted as the attitude $a_{i,t}$ for the subject $i$'s estimate at time step $t$. In the decentralized network condition and the decentralized network condition, subjects got to see the arithmetic average estimate of its neighbors $N_i$ in a network $G$ ($\bar{m}_{i,t} = \frac{1}{(N_J)} \cdot \sum_{j \in J_i} a_{j,t}$, where $J_{i,t} = N_i$). In the decentralized network condition, the network was a random graph where each subject had the same number of neighbors (4 neighbors). In the centralized condition, the network was a star graph where everyone was only connected to one central subject. In both the network conditions, after they saw the average message ($\bar{m}_{i,t}$), they were asked again to provide their updated estimates for the same question ($a_{i,t+1}$ for all $i$). In the control condition, the subjects also underwent three rounds of estimation tasks, except no message was presented to them in between. At the end of the experiment, the subjects received monetary rewards based on the accuracy of their estimates.

To test the hypothesis that the social influence can improvement of WOC effect due to the role of confidence described above, they simulated with an ABM where within the attitude update function, the social influence strength parameter $\alpha_i$ is positively correlated with one's error, as measure by $|\alpha_i - a^*|$ (or, equivalently, the "self-weight" $(1 - \alpha_i)$ is negatively correlated with the error)[5].

**Attitude update function.** For each subject $i$, the experimental design presented the subject with the arithmetic average of his neighbors' messages, rather than each individual neighbor's message. Because of this, they formulated the attitude update function accordingly to correspond to the experimental design. The attitude update function thus needed to be modified slightly from Equation (*2*) to

$$\Delta a_{i,t} = \alpha_{i,t} \cdot f_{update}\big(a_{i,t}, \ \bar{m}_{i,t}\big), \tag{29}$$

where $\bar{m}_{i,t} = \frac{1}{(N_J)} \cdot \sum_{j \in J_{i,t}} m_{j,t} = \frac{1}{(N_J)} \cdot \sum_{j \in J_{i,t}} a_{j,t}$ (the arithmetic average) and $J_{i,t} = N_i$.

They further assumed that the attitude change can be expressed by

$$\Delta a_{i,t} = \alpha_i \cdot \big(\bar{m}_{i,t} - a_{i,t}\big). \tag{30}$$

This formulation was as a special case of DeGroot (1974)' model (Equation (*12*)), in which $p_{i,j} = \frac{1}{N_J}$. Indeed, they also assumed $\alpha_i \in [0,1]$ according to DeGroot (1974)'s model.

Based on their hypothesis about the role of $\alpha_i$, they assumed $\alpha_i$ to be a function of $|\bar{m}_{i,t} - a_{i,t}|$, i.e.,

$$\alpha_i = f_\alpha\big(|\bar{m}_{i,t} - a_{i,t}|\big) \tag{31}$$

and the correlation between $|\bar{m}_{i,t} - a_{i,t}|$ and $\alpha_i$, i.e., $r(\alpha_i, |\bar{m}_{i,t} - a_{i,t}|)$, was manipulated between 0 and 1.

- ***Aggregation Function.*** No aggregation function because the messages are averaged into $\bar{m}_i$ when presented to the subject $i$.

- **Assimilation Force.** $asm\big(a_{i,t}, \bar{m}_{i,t}\big) = \big(\bar{m}_{i,t} - a_{i,t}\big)$.

---

[5] In their formulation, they used $a_{i,t+1} = (1 - \alpha_i) \cdot a_{i,t} + \alpha_i \cdot \big(\bar{m}_{i,t} - a_{i,t}\big)$, and referred $(1 - \alpha_i)$ as "self-weight", the weight one gives to himself relative to the message he receives.

- Reinforcement Force. None.

- Similarity Bias. None.

- Repulsion Force. None.

- **Other Assumptions.** (1) $\alpha_i = f_\alpha\big(|\bar{m}_{i,t} - a_{i,t}|\big) \in [0,1]$.

**Selection Function.** For the centralized and decentralized conditions, $J_{i,t} = N_i = \{$all neighbors of the subject $i$ in the network $G\}$.

- ***Number of Sources.*** $N_J = |N_i|$. Note that the subject only receives one single message, the average message across $N_J$ sources, at a time.

**Message Function.** $\bar{m}_{i,t} = f_{message}\big(\{a_{j,t}\}_{j \in N_J}\big) = \bar{a}_{j,t} = \frac{1}{(N_J)} \cdot \sum_{j \in J_{i,t}} a_{j,t}$. Because in the experimental design, the message of the neighbors is simply the average of their estimates.

**Simulation results.** In the decentralized network condition (Figure S9, S10, in Becker et al., 2017), they show they the attitude distribution $f_{A,t}(a)$ converge to a consensus. If $r\big(\alpha_i, |\bar{m}_{i,t} - a_{i,t}|\big) > 0$, the median of the attitude distribution moves towards the ground truth. Otherwise, if $r\big(\alpha_i, |\bar{m}_{i,t} - a_{i,t}|\big) < 0$, the median moves away from the ground truth. In the centralized network condition, the attitude of the central node (corresponds to the central subject in the human experiment; $a_{central,t}$) plays a vital role. If $a_{central,t}$ is in the same direction as the ground truth relative to the initial distribution median, i.e., $(a_{central,t} - \text{median}(f_{A,t=1}(a))) \cdot (a_{central,t} - a^*) > 0$, then social influence will improve the WOC effect, i.e., $\text{median}(f_{A,t=1}(a)))$ will get closer to $a^*$.

**Empirical Evidence.** The empirical evidence was in line with the simulation results. Consistent with simulations, they showed that in the decentralized network condition, the median of the group estimate did decrease after receiving the neighbors' message. Critically, they found that the social influence strength is positively correlated with one's accuracy, i.e., $r\big(\alpha_i, |\bar{m}_{i,t} - a_{i,t}|\big) > 0$. This supported their hypothesis that the improvement of the WOC effect under social influence was driven by the correlation between one's accuracy and one's willingness to change his attitude. The results in the centralized network condition were also consistent with the simulations

**Remarks.** In Becker et al. (2017)'s study, they found that the social strength parameter $\alpha_i$ is a function of the accuracy of the subject's attitude. To my understanding, this factor has never been modeled in any deductive ABM. Nonetheless, this factor seems to be a driving force in group opinion dynamics, which leads to the WOC effect.

One limitation of this study is that the message was shown as an average of attitudes ($\overline{m}_{i,t}$). Indeed, in the experimental design, the message was shown in the phrase "Average of other players: [XXX]". Subjects may regard a message differently depending on whether the message is from an individual or from a group average.

Another limitation is that the subjects were incentivized to provide accurate attitude. The monetary compensation they received depended on how accurate their attitudes were. Without this incentive, people may lose the motivation to update their attitudes based on their confidence.

### Becker et al. (2019)

Related to the WOC effect, another hypothesis that was tested with ABM is the *group polarization hypothesis*. According to this hypothesis, attitude regarding political belief usually involves a *partisan bias*. Studies have shown that, for a partisan topic, people with stronger *partisan bias* are less likely to update their attitudes upon receiving messages from other people (Sunstein, 2009). Note that although the *partisan bias* seems similar to the *polarization factor* included in the deductive ABMs (e.g., Equation (*11*)), there exists a critical and subtle difference. Let me explain this with a concrete example. While Democrats in general have the partisan bias to underestimate the number of unauthorized immigrants in the U.S., a Democrat with a stronger partisan bias tend to underestimate more. In contrast, Republicans in general have a partisan bias to overestimate the number of unauthorized immigrants. A Democrat that underestimates the number greatly is regarded as having a stronger partisan bias, while a Republican that underestimates the number with the same amount is conversely regarded as having a less partisan bias because it does not align with the attitude of the party.

They further predicted that, when people only interacted with like-minded people ("partisan crowd"), if the partisan bias indeed controls how receptive a person is to the social influence, then the WOC effect should be undermined. That is, the group median attitude would be farther away from the true $a^*$, and the group median attitude would become more extreme after social influences. Formally, $|\text{median}(f_{A,t+1}(a)) - a^*| > |\text{median}(f_{A,t}(a)) - a^*|$ and $\text{median}(f_{A,t+1}(a)))| > |\text{median}(f_{A,t}(a)))|$. This is due to the fact that those with more extreme attitudes (which tend to be more erroneous) will impact others' attitudes than the other way around. To demonstrate this effect, they formulated an ABM based on the polarization hypothesis, and simulated the aforementioned phenomena.

Conversely, they also predicted that, if the WOC effect is robust to the partisan bias, then the updated attitude distribution should resemble to Becker et al. (2017)'s findings. They termed this phenomenon as the *wisdom of partisan crowds*.

**Experimental Designs.** They recruited 1120 subjects, where 560 self-identified as Republican and the other 560 as Democrats. Within each party, subjects were randomly assigned into 16 separate groups. Each group had 35 subjects who interacted with each other. Each group belonged to one of the two conditions, a homogenous network condition (12 groups) and a control condition (4 groups). Across all three conditions, all subjects went through three rounds of estimation tasks ($t = 1, 2, 3$) for several different topics. In each estimation task, each subject provides an estimate about a question that was known to be partisan (e.g., the number of unauthorized immigrates in the US), denoted as the attitude $a_{i,t}$ for the subject $i$'s estimate at time step $t$. In the homogenous network condition, subjects got to see the arithmetic average estimate of its like-minded neighbors $N_i$ (size = 4) in a random network $G$ ($\bar{m}_{i,t} = \frac{1}{(N_J)} \cdot \sum_{j \in J_i} a_{j,t}$, where $J_{i,t} = N_i$). After they saw the average message, they were asked again to provide their updated estimates for the same question. In the control condition, the subjects also underwent three rounds of estimation tasks, except no message was presented to them in between. At the end of the experiment, the subjects received monetary rewards based on the accuracy of their estimates.

To test the hypothesis that the social influences among like-minded people lead to group polarization, they formulated an ABM where within the attitude update function, there exists a partisan bias that modulates the social influence strength parameter $\alpha_i$ (or, equivalently, the "self-weight" $(1 - \alpha_i)$ [6]), such that it is negatively correlated with one's partisan bias (defined below).

**Attitude update function.** Because they adopted a very similar experimental design as in Becker et al. (2017), the attitude change can also be expressed by

$$\Delta a_{i,t} = \alpha_i \cdot \left( \bar{m}_{i,t} - a_{i,t} \right). \tag{32}$$

Based on their hypothesis about partisan bias, they further assumed

$$\alpha_i = \sigma \left( a_{i,t} \cdot b_i + e_i \right), \tag{33}$$

where $\sigma$ is the sigmoid function, $b_i = +1$ if the agent belongs to a class that tend to overestimate (e.g., Republicans estimating the number of unauthorized immigrants), and $b_i = -1$ if the agent belongs to a class that tend to underestimate (e.g., Democrats estimating the number of unauthorized immigrants), and $e_i \sim N(0, \varepsilon)$ is the random error (they tested $\varepsilon = 0, 1, 5$). In the simulation, they manipulated the proportion of agents with the same partisan bias in the network $G$.

---

[6] In their formulation, they used $a_{i,t+1} = (1 - \alpha_i) \cdot a_{i,t} + \alpha_i \cdot \left( \bar{m}_{i,t} - a_{i,t} \right)$, and referred $(1 - \alpha_i)$ as "self-weight", the weight one gives to himself relative to the message he receives.

- *Aggregation Function.* No aggregation function because the messages are averaged into $\bar{m}_i$ when presented to the subject $i$.

- **Assimilation Force.** $asm\big(a_{i,t}, \bar{m}_{i,t}\big) = \big(\bar{m}_{i,t} - a_{i,t}\big)$.

- Reinforcement Force. None.

- Similarity Bias. None.

- Repulsion Force. None.

- **Other Assumptions.** (1) $\alpha_i = \sigma\big(a_{i,t} \cdot b_i + e_i\big), \in [0,1]$.

**Selection Function.** When agents are connected in a network, $J_{i,t} = N_i = \{$all neighbors of the subject $i$ in the network $G\}$.

- *Number of Sources.* $N_J = |N_i|$. Note that the subject only receives one single message, the average message across $N_J$ sources, at a time.

**Message Function.** $\bar{m}_{i,t} = f_{message}\big(\{a_{j,t}\}_{j \in N_J}\big) = \bar{a}_{j,t} = \frac{1}{(N_J)} \cdot \sum_{j \in J_{i,t}} a_{j,t}$. Because in the experimental design, the message of the neighbors is simply the average of their estimates.

**Simulation results.** In the network condition (Figure S4, in Becker et al., 2019), they showed that when there are more like-minded agents (having the same partisan bias) in a network, the more polarized a group's median attitude becomes and the farther away it is from the ground truth. Formally, $|\text{median}(f_{A,t+1}(a))) - a^*| > |\text{median}(f_{A,t}(a))) - a^*|$ and $|\text{median}(f_{A,t+1}(a)))| > \text{median}(f_{A,t}(a)))|$.

**Empirical Evidence.** The empirical evidence was *against* the simulation results. Contrary to the simulations, they showed that in the homogenous network condition, the median of the group estimate decrease after receiving the neighbors' message. In other words, they decrease their partisanship after interacting with like-minded people. This contradicted with the predictions from the group polarization hypothesis. In conclusion, this supported the hypothesis that the WOC effect is robust to the partisanship of a topic.

**Remarks.** In Becker et al. (2019)'s study, they tested the group polarization hypothesis which assumes that the social strength parameter $\alpha_i$ is a function of the subject's partisan bias. They formulated this hypothesis and tested it with human experiments. The experiment results did not support the group polarization hypothesis.

The two limitations identified above in Becker et al. (2017) also applied to this study. That is, 1) the message was shown as an average of attitudes, and 2) the subjects were incentivized to provide accurate attitude

## Data-driven Modeling

Aside from hypothesis testing, another application of ABMs is to directly induce attitude update rules based on observed empirical data. For example, researchers can use ABM to quantify specific parameters in the attitude update function (e.g., the strength of the social influence $\alpha_i$), and then propose mechanisms to explain the observed patterns in the parameters. For instance, Jayles et al. (2017) found that as the difference between a person's initial attitude and the message he receive ($\left| m_{j,t} - a_{i,t} \right|$) increased, the strength of social influence ($\alpha_i$) of a message on him also got larger (with a linear cusp). This is opposite to the *similarity bias* assumed in the deductive ABMs. (Becker et al., 2017, 2019; Jayles et al., 2017). I will elaborate the studies in this literature below, including their experimental designs, the ABM formulations, the parameters of interest, and model-fitting results.

### *Jayles et al (2017)*

Related to the study on social influence on the WOC effect, Jayles et al (2017) studied whether and how social influence between people can improve their accuracy if they influence one another in a sequential order - only one subject receive social influence at a time. They used an ABM to quantify the social influence strength term and explained the individual differences in it.

**Experimental Design.** 360 subjects were grouped into 18 groups, where each group had 20 subjects. Similar to the design in Becker et al. (2017) and Becker et al. (2019), each subject was asked to provide estimates $x_{i,t}$ on some questions (e.g., "How many balls do you think are in this jar?"). However, in Jayles et al (2017)'s design, subjects were doing the task sequentially, rather than in parallel as in Becker et al. (2017) and Becker et al. (2019). At each time step $i$, one unique subject $i$ provides its initial estimate $x_{i,initial}$, he then receives the average of the previous $\tau$ subjects' estimates ($\tau \in \{1,3,5\}$) as message $\bar{m}_i$. Formally, $\bar{m}_i = \frac{1}{(N_J)} \cdot \sum_{j \in J_i} x_{j,initial}$, where $J_i$ is the set of the previous $\tau$ subjects in the previous $\tau$ time steps[7]. After the subject $i$ receive the average message $\bar{m}_i$, the subject is then asked to provide his estimate again, denoted as $x_{i,updated}$. Because there are 20 subjects, this process repeated 20 times until everyone gave their estimates. At the end of the experiment, the subjects received monetary rewards based on the accuracy of their estimates.

When they formulated the ABM, they first normalized the estimates by the ground true answer to the question($T$) and log-transform the estimate. Formally, $a = log(x/T)$. I denote this normalized term as attitude $a_i$ to be consistent with the notations used throughout the text.

**Attitude update function.** The attitude update function is expressed as

$$\Delta a_i = a_{i,updated} - a_{i,initial} = \alpha_i \cdot \left( \bar{m}_i - a_{i,initial} \right). \tag{34}$$

While this is similar to the DeGroot (1974)'s classic averaging model (Equation (36)), there are some critical differences. First, the social influence strength parameter $\alpha_i$ is unbounded, i.e., $\alpha_i \in \mathbb{R}$. This allows one to identify those who overreact to the message ($\alpha_i >$

---

[7] In the actual experiment, they conducted the experiment in Japan and France separately. In Japan, the arithmetic mean of the estimate was shown as the message, whereas in France, the geometric mean was shown instead. For the sake of simplicity. I used arithmetic mean in the formulation. In addition, they applied special treatment for the first $\tau - 1$ subjects.

1) and those who move away from the direction of the message ($\alpha_i < 1$). Second, $\alpha_i$ is estimated from the data rather than a fixed constant.

- ***Aggregation Function.*** No aggregation function because the messages are averaged into $\bar{m}_i$ when presented to the subject $i$.

- **Assimilation Force.** $asm(a_{i,t}, m_i) = (\bar{m}_i - a_{i,initial})$.

- Reinforcement Force. None.

- ***Similarity Bias.*** Allowed to be modeled because $\alpha_i$ is estimated from the data and can be modeled with a similarity bias (see below).

- ***Repulsion Force.*** Allowed to be modeled because $\alpha_i$ is unbounded and estimated from the data.

- Other Assumptions. None.

**Selection Function.** The $\tau$ subjects that did the task prior to the subject $i$.

- ***Number of Sources.*** $N_J = \tau$. Note that the subject only receives one average message .

**Message Function.** $\bar{m}_i = \frac{1}{(N_J)} \cdot \sum_{j \in J_{i,t}} x_{j,initial}$. Because in the experimental design, the is simply the average of their estimates.

**Parameters of Interest.** The parameter of interest is $\alpha_i$, the weight the subject $i$ gives to the messages $\bar{m}_i$. According to Equation (34), the value of $\alpha_i$ can be estimated by

$$\hat{\alpha}_i = \frac{\Delta a_i}{(\bar{m}_i - a_{i,initial})} \tag{35}$$

**Model-Fitting Results.** The estimated $\hat{\alpha}_i$ across subjects is a function of the distance $|\bar{m}_i - a_{i,initial}|$ as shown in Figure 1. Figure 1A. shows that the larger the $|\bar{m}_i - a_{i,initial}|$ is, the larger $\alpha_i$ is (with a linear cusp relationship). Figure 1B. shows the mechanism underlying this linear cusp relationship. The three colors represent three main types of responses under social influence: 1) *compromisers* who compromise towards the message ($0 \leq$

$\hat{\alpha}_i \leq 1$), 2) *keepers* who keep the initial attitude and ignore the message ($\hat{\alpha}_i = 0$), and 3) *adopters* who adopt the message and ignore the initial attitude ($\hat{\alpha}_i = 1$). As shown in Figure 1B, when $|\bar{m}_i - a_{i,initial}|$ increases up to a certain cutoff value, the proportion of those who keep decreases, and the proportion of those who compromise increases. This drives the linear cusp relationship as shown in Figure 1A. They also proposed a mathematical model to fit this pattern in the data, which I will not include the model here as it is convoluted.

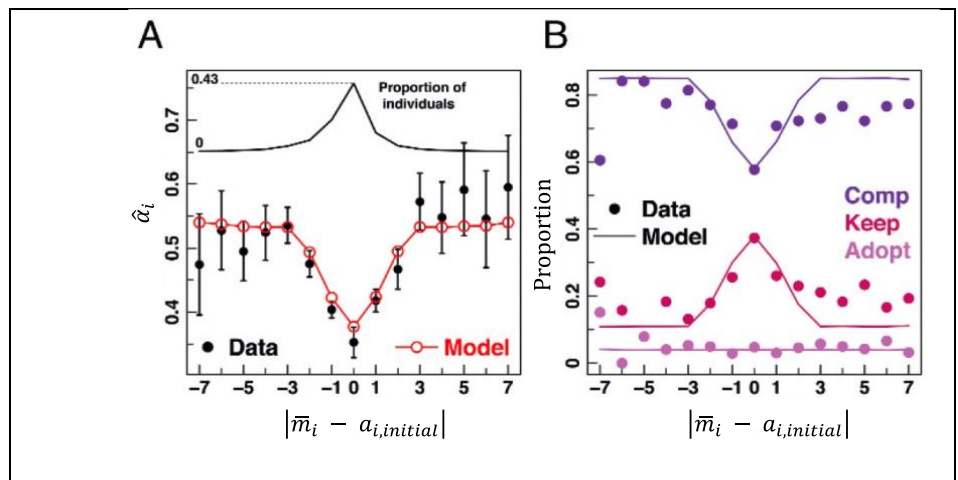

Figure 1. (A) The relationship between the estimated $\hat{\alpha}_i$ and the distance $|\bar{m}_i - a_{i,initial}|$. The red line is based on their model prediction, and the black line is the proportion of subjects with the specific value of $|\bar{m}_i - a_{i,initial}|$. (b) The proportion of subjects who are keepers ($\hat{\alpha}_i = 0$), compromiser ($0 \leq \hat{\alpha}_i \leq 1$), or adaptors ($\hat{\alpha}_i = 1$), as a function of $|\bar{m}_i - a_{i,initial}|$. Modified from Figure 2 in Jayles et al (2017)

**Remarks.** Jayles et al (2017) adopted an ABM to estimate the social influence strength parameter $\hat{\alpha}_i$ and studied the individual differences in it. They found that $\hat{\alpha}_i$ changed as a function of the distance $|\bar{m}_i - a_{i,initial}|$. Contrary to the similarity bias assumed in deductive ABMs, they found that $\hat{\alpha}_i$ *increases* as a function of $|\bar{m}_i - a_{i,initial}|$, which is driven by that some keepers become compromisers when $|\bar{m}_i - a_{i,initial}|$ gets larger. To my knowledge, the

positive correlation between $\left|\overline{m}_i - a_{i,initial}\right|$ and $\hat{\alpha}_i$ has never been modeled in any deductive ABM.

The two limitations identified above in Becker et al. (2017) and Becker et al. (2019) also applied to this study. That is, 1) the message was shown as an average of attitudes, and 2) the subjects were incentivized to provide accurate attitude.

**Frigo (2022)**

Frigo (2022) suggested a mechanism for why people sustain false belief even after receiving information that contradict with their initial belief. They used the *category-learning task* to manipulate and measure subjects' attitude. They that found when there were two messages presented to a subject, if one message closely aligned with the subject's attitude, the subject would give full weight to this message and completely ignore the other message. They proposed a *heuristic evidence weighting* (HEW) model to describe the pattern in the data.

**Experimental Design.** In the experiment[8], each of the 66 subjects underwent a *category-learning task*. In the category-learning task, subjects were asked to classify some artificial fruits into two classes – whether eaten after cooked or eaten in raw (the label $y \in \{raw = 0, cooked = 1\}$). The shape of the fruit varied in a one-dimensional feature space $X = [0,300] \in \mathbb{Z}$. There were 301 fruits in total, each took a unique value $x$ in the feature space. The attitude $a_{i,t}$ of any subject $i$ at any given time $t$ is then defined as the decision boundary he adopts for the binary classification, i.e., $y = 0$ if $x \geq a_{i,t}$, and otherwise $y = 1$. In the first phase of the task, each subject underwent a *supervised learning phase*, where there was an "oracle" that with attitude $a^* = 150$ communicating its attitude with the subject. In this first phase, the oracle's attitude was communicated through a series of 20 tuples $\{(x_k, y_k^*)\}_{k=1,\ldots 20}$, where $y_k^*$ is the label provided by the oracle. In the second phase, the subject initial attitude $a_{i,t=1}$ was then measured in an *initial boundary measurement phase*. Because the effect of the supervised learning phase, the initial attitude should be roughly equal 150 for all subjects. The third phase is the *learning from sources phase*, the subjects received messages from two "artificial agents" (denoted as $m$ and $n$), who have fixed attitudes and do not update their attitudes over the time. Let denote their decision boundaries as $a_m$ and $a_n$. The messages were communicated in the same tuples form, i.e., $\left\{(x_k, y_{m,k}, y_{n,k})\right\}_{k=1,\dots301}$ for the agent $m$ and the agent $n$ over the entire feature space $X = [0,300]$. In the fourth phase, each subject then underwent a *final boundary measurement phase*, where the updated attitude $a_{i,t=2}$ was measured.

---

8 While Frigo (2022) designed multiple experiments, I will focus on the experiment two because it was where the ABM was induced to model the empirical data.

In the experiment, they fixed the attitude of an agent $n$ to either 135 or 165 (15 units away from $a^* = 150$) and varied the attitude of another agent $m$. They were interested in how $\Delta a_{i,t}$ changed as a function of $|a_m - a_{i,t}|$.

**Attitude update function.** The attitude update function is expressed as

$$\Delta a_{i,t} = \sum_{j \in J_{i,t}} p_{ij}\left(a_{j,t} - a_{i,t}\right) \tag{36}$$

, where $p_{ij}$ is the weight that the subject gives to each of the two artificial sources. This can be regarded as a variant of the DeGroot (1974)'s classic averaging model (Equation (36)). This expression differs from the original form in that 1) the weight $p_{ij}$ is not a constant, but rather a parameter to be estimated from the data, and 2) the social influence strength is assumed to be 1 ($\alpha_i = 1$). The assumption of $\alpha_i = 1$ is equivalent to assuming that the subject gives no weight to his initial attitude when exposed to others' messages.

- **Aggregation Function.** A weighted average function. Formally, $agg\left(\left\{g\left(a_{i,t}, m_{j,t}\right)\right\}_{j \in J_{i,t}}\right) = \sum_{j \in J_{i,t}} p_{ij} g\left(a_{i,t}, m_{j,t}\right)$ , where $p_{ij} \in [0,1]$ and $\sum_{j \in J_{i,t}} p_{ij} = 1$.

- **Assimilation Force.** $asm\left(a_{i,t}, m_{j,t}\right) = \left(m_{j,t} - a_{i,t}\right)$.

- Reinforcement Force. None.

- **_Similarity Bias._** Allowed to be modeled because $p_{ij}$ is estimated and can be modeled with a similarity bias term (see below).

- **_Repulsion Force._** None. Note that the constraint $p_{ij} \in [0,1]$ does not allow a potential repulsion force (which requires $p_{ij} < 0$).

- **_Other Assumptions._** (1) $\alpha_i = 1$. Note that this is a strong assumption because is equivalent to assuming that the subject gives no weight to his initial attitude when exposed to others' messages

**Selection Function.** The two sources that provide messages to the subject are artificial agents that have fixed attitudes and do not update their attitudes under social influences. $J_{i,t} = \{m, n\}$.

- **_Number of Sources._** $N_J = 2$. Note that the subject always receives two messages at a time.

**Message Function.** $m_{j,t} = f_{message}(a_{j,t}) = a_{j,t}$. Note that the decision boundary $a_{j,t}$ of the two artificial agents $m$ and $n$ are conveyed in the form of tuples $\{(x_k, y_{m,k})\}_{k=1,\dots 301}$ and $\{(x_k, y_{n,k})\}_{k=1,\dots 301}$.

**Parameters of Interest.** The parameter of interest is $p_{i,m}$, the weight the subject $i$ gives to $m$, the agent whose attitude is manipulated across subjects. Its attitude is now denoted as $a_{m,i}$, and the subscript $i$ entails that its value varies across subjects. According to Equation (36), the value of $p_{i,m}$ can then be estimated by

$$\hat{p}_{i,m} = \frac{(a_{i,t=2} - a_n)}{(a_{m,i} - a_n)} \tag{37}$$

**Model-Fitting Results.** The estimated $\hat{p}_{i,m}$ across different various values of $a_m$ can be fitted with the following *heuristic evidence weighting* (HEW) model,

$$\hat{p}_{i,m} = 1 - \frac{\max\left(|a_{m,i} - a_{i,t=1}| - \alpha, 0\right)}{\max\left(|a_{m,i} - a_{i,t=1}| - \alpha, 0\right) + \beta} \tag{38}$$

where $\alpha$ and $\beta$ control the shape of the curve. Note that as $|a_{m,i} - a_{i,t=1}|$ gets larger,

the weight $\hat{p}_{i,m}$ decays in a non-linear way (Figure 2.)

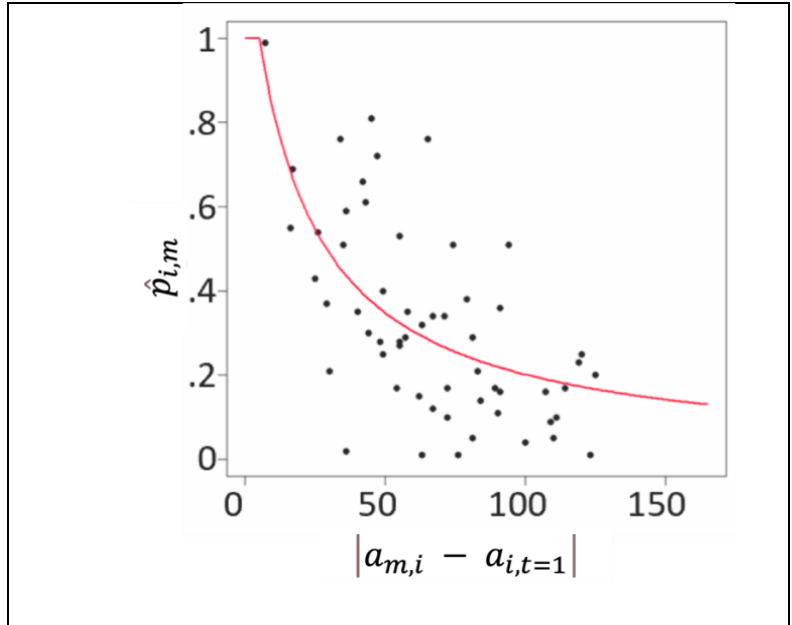

Figure 2. The heuristic evidence weighting (HEW) model (red line) estimated based on the experiment data. Each dot represents a subject's distance to agent $m$'s attitude $a_{m,i}$ ($|a_{m,i} - a_{i,t=1}|$) and the estimated weight for that agent $m$ ($\hat{p}_{i,m}$). Modified from Figure 5 in Frigo (2022).

**Remarks.** Frigo (2022) found that there is a monotonically decaying relationship between a subject's distance to artificial agent $m$'s attitude $a_{m,i}$ ($|a_{m,i} - a_{i,t=1}|$) and the estimated weight for that artificial agent ($\hat{p}_{i,m}$). Note that this is conceptually similar to the similarity bias term in Lorenz et al. (2021)'s deductive ABM (e.g., compared Equation (*24*) and Equation (38)), except that HEW requires two messages being presented simultaneously.

One key breakthrough of their work is that they found that when one of the two sources (denoted as $m$) lies within $\pm5$ units of the agent's initial attitude $a_{i,t=1}$, this source will have $\hat{p}_{i,m} = 1$. This can be observed from Equation (38) if $\alpha \leq 5$. That is, the subject will give all weight to agent $m$'s attitude $a_{m,i}$ and completely ignore the other agent $n$'s attitude $a_{n,i}$. This prediction of the HEW model is supported by the empirical data in their experiment two and three.

To my knowledge, this work is the first to show that a subject's similarity bias towards one source $j$ also depends on its distance to a separate source $k$. Formally, they showed that $sim(a_{i,t}, m_{j,t})$ is not only a function of distance$(a_{i,t}, m_{j,t})$, but also a function of the distance$(a_{i,t}, m_{k,t})$. This dependency has never been proposed in any deductive ABM.

One critical limitation of the HEW model is that it requires exact two sources at a time ($N_J = 2$). It does not define what should happen in the case of single source ($N_J = 1$) or multiple sources ($N_J > 2$). Another limitation of this ABM is that is assumes the social influence strength $\alpha_i = 1$, which is a strong assumption because it assumes the subject to completely discard what he believes prior to the social influence. They did not justify this assumption empirically.

**Discussion on Deductive ABMs**

In the section above, I delineate two major ways of leveraging ABMs to study attitude update rules from human experiments, 1) hypothesis testing and 2) data-driven modeling. An example that benefits from hypothesis testing with ABMs is the wisdom of crowds effect. Empirical results suggest that the degree to which an individual will change his attitude due to social influence ($\alpha_i$) depends on how accurate his initial attitude is, and this effect underpins the improvement of collective accuracy after social interaction (Becker et al., 2017, 2019). Aside from hypothesis testing, studies using the data-driven modeling approach have bootstrapped attitude update functions using ABMs (Frigo, 2022; Jayles et al., 2017). Jayles et al. (2017) found a linear cusp relationship between the social influence strength ($\alpha_i$) and the difference between the social information and the person's initial attitude. Frigo (2022) found that the similarity bias of a subject towards an agent depends not only on the distance between their attitudes, but also depends on another agent's attitude. To my knowledge, both these relationships have never been included in any deductive ABMs.

There are several major limitations existing work on deductive ABMs. First, the messages are shown as a group average ($\bar{m}_i$) rather than the attitude of each individual ($M_{i,t} = \{m_{j,t}|j \in J_{i,t}\}$) (Becker et al., 2017, 2019; Jayles et al., 2017). In most real-world scenario, when a person receives multiple messages simultaneously, e.g., a Twitter user viewing different comments about a topic, the person is able to see the content of each individual message, rather than an average of the messages. Moreover, it has been shown that the update of a person's attitude upon receiving two messages is distinct from the update of attitude upon receiving one average message (Frigo, 2022). Therefore, it remains unclear as to how much their findings generalize to real-world opinion dynamics. Second, all aforementioned studies have implicit constraints in their ABM formulations, and these constraints are not empirically justified. For example, all deductive ABMs above assume the assimilation force but not the reinforcement force, which is a critical factor that can lead to bi-polarization and extremization (Lorenz et al., 2011). Similarly, Becker et al (2017, 2019) constrained the social influence strength $\alpha_i$ to be bounded in $[0,1]$. The constraint disallows the exploration of those who overreact to the message or those who get repulsed away from the message, who collectively account for up to 20% of people in an empirical study (Jayles et al., 2017). Frigo (2022) even assumed $\alpha_i = 1$, a strong assumption that is not justified either.

## Conclusions

In this review paper, I survey deductive ABMs and inductive ABMs regarding attitude change under social influences. I propose a general formulation to unify ABMs in studies from a wide variety of disciplines. With the aid of this formulation, I show that all ABMs can be viewed as special cases of this general formulation. This angel highlights the implicit assumptions imposed by each ABM, e.g., the range of the social influence strength, the number of messages at a time. In addition, I propose the contrast between deductive ABMs and inductive ABMs, where the former refers to work in social simulations and the latter refers to ABMs used in empirical studies. I show that deductive ABMs have not yet been verified rigorously, and empirical studies with inductive ABMs reveal factors that are not assumed in any deductive ABM. This highlights a huge unexplored regime for those who want to improve and validate/falsify existing deductive ABMs. Besides, I also mention the limitations of existing work using inductive ABMs, where the experimental design or/and the ABM formulations impose unrealistic constraints.

In the following section, I would like to pinpoint several potential future directions for the research on ABMs. First, in social simulations, the effects of different choices of the selection function, the message function, and the aggregation function are underexplored. For example, for the selection function, one can explore whether there exists a particular recommendation algorithm for social media that can prevent the formulation of echo chambers or extremization. To date, the only simulation for this purpose is with a toy example where there are only 10 agents interacting with one another (Frigo, 2022). For the message function, one can simulate a system where some agents deliberately exaggerate messages (e.g., some media outlets). Another underexplored component is the aggregation function. As for the aggregation function, to my surprise, many deductive ABMs only allow one message at a time. This is an unrealistic assumption because most social interaction involves more than two persons at a time (e.g., social media). Furthermore, for the inductive ABMs that allow more than one simultaneous message, the aggregation function is often assumed to be a simple average or summation function. According to empirical studies, the aggregation function should be more complicated. For example, when a message is closely aligned with one's attitude, then it can get all weight and leave zero weight for other messages (Frigo, 2022). Therefore, it is worth exploring the effect of different aggregation functions.

Second, deductive ABMs in social simulations should incorporate evidence found in inductive ABMs in human experiment studies (Becker et al., 2017, 2019; Frigo, 2022; Jayles et al., 2017). Deductive ABMs can include extra factors that have empirical grounds, e.g., the dependency between similarity biases towards two agents (Frigo, 2022). Similarly, to validate or falsify the formulation of inductive ABMs, modelers can compare their simulation results against empirical results in deductive ABMs. For example, the role of polarization is falsified in partisan political topic (Becker et al., 2019). In addition, while some deductive ABM studies attempt to compare various ABMs against macro-level attitude distribution dataset and see which ABM fits the best, the comparison is not done in a rigorous and systematic way (Lorenz et al., 2021). The field of social simulation would benefit from a common benchmark dataset where all deductive ABMs can be evaluated upon.

Third, human experiments can be carefully designed to better connect deductive ABMs with inductive ABMs. While some empirical studies on inductive ABMs shed light on how people update their attitude, the scenario is usually overly simplified, e.g., by showing the average message (Becker et al., 2017, 2019; Jayles et al., 2017), or by limiting number of messages to two (Frigo, 2022). One straightforward yet critical modification to Becker et al. and Jayles et al (2017, 2019; 2017)'s design is to allow the each neighbor's message to be shown to the subject. This allows one to explore the aggregation function that people actually use.

A better understanding of ABMs for attitude update has a wide application. For instance, if we have a reliable ABM that predicts how human change their belief, it enables possible intervention or manipulation of undesired macro-level attitude distribution like bipolarization or extremization. For example, one may be able to devise better recommendation algorithms for social media to prioritize selected messages to users to mitigate echo chambers (for an attempt, see the appendix in Frigo, 2022). Apart from opinion dynamics, *influence limitation* is another possible application of ABMs. The problem of influence limitation refers to the setting where we want to limit the spread of misinformation in a social network by launching a spread of corrective message, it requires an ABM that defines how a person should update their beliefs while receiving opposing messages (Budak et al., 2011). Within the field of influence limitation, the common ABMs in use are unrealistically naïve and without empirical grounds. With a more realistic ABM, the solutions to influence limitation should be more applicable to real-world scenarios. Another relevant application is in viral marketing, where competing or complementary products are adopted by users, and the marketer wants to leverage social influences to benefit the efficacy of a marketing campaign (Lu et al., 2015).

## References

Akers, R. L., Krohn, M. D., Lanza-Kaduce, L., & Radosevich, M. (1995). Social learning and deviant behavior: A specific test of a general theory. *Contemporary Masters in Criminology*, 187–214.

Anderson, N. H. (1971). Integration theory and attitude change. *Psychological Review*, *78*(3), 171.

Banerjee, S., Jenamani, M., & Pratihar, D. K. (2020). A survey on influence maximization in a social network. *Knowledge and Information Systems*, *62*(9), 3417–3455.

Banisch, S., & Olbrich, E. (2019). Opinion polarization by learning from social feedback. *The Journal of Mathematical Sociology*, *43*(2), 76–103.

Baumann, F., Lorenz-Spreen, P., Sokolov, I. M., & Starnini, M. (2020). Modeling echo chambers and polarization dynamics in social networks. *Physical Review Letters*, *124*(4), 048301.

Becker, J., Brackbill, D., & Centola, D. (2017). Network dynamics of social influence in the wisdom of crowds. *Proceedings of the National Academy of Sciences*, *114*(26), E5070–E5076.

Becker, J., Porter, E., & Centola, D. (2019). The wisdom of partisan crowds. *Proceedings of the National Academy of Sciences*, *116*(22), 10717–10722.

Budak, C., Agrawal, D., & El Abbadi, A. (2011). *Limiting the spread of misinformation in social networks*. 665–674.

Centola, D., Becker, J., Brackbill, D., & Baronchelli, A. (2018). Experimental evidence for tipping points in social convention. *Science*, *360*(6393), 1116–1119.

Deffuant, G., Amblard, F., Weisbuch, G., & Faure, T. (2002). How can extremism prevail? A study based on the relative agreement interaction model. *Journal of Artificial Societies and Social Simulation*, *5*(4).

Deffuant, G., Neau, D., Amblard, F., & Weisbuch, G. (2000). Mixing beliefs among interacting agents. *Advances in Complex Systems*, *3*(01n04), 87–98.

DeGroot, M. H. (1974). Reaching a consensus. *Journal of the American Statistical Association*, *69*(345), 118–121.

Dodds, P. S., & Watts, D. J. (2005). A generalized model of social and biological contagion. *Journal of Theoretical Biology*, *232*(4), 587–604.

Doob, L. W. (1947). The behavior of attitudes. *Psychological Review*, *54*(3), 135.

Festinger, L. (1962). *A theory of cognitive dissonance* (Vol. 2). Stanford university press.

Flache, A., Mäs, M., Feliciani, T., Chattoe-Brown, E., Deffuant, G., Huet, S., & Lorenz, J. (2017). Models of social influence: Towards the next frontiers. *Journal of Artificial Societies and Social Simulation*, *20*(4).

Frigo, V. (2022). *An Examination of Non-Normative Belief Updating Behavior in Humans (Why Is It So Hard to Change Minds?)*. UNIVERSITY OF WISCONSIN-MADISON.

Galton, F. (1907). Vox Populi. *Nature*, *75*(1949), 450–451. https://doi.org/10.1038/075450a0

Hegselmann, R., & Krause, U. (2002). Opinion dynamics and bounded confidence models, analysis, and simulation. *Journal of Artificial Societies and Social Simulation*, *5*(3).

Holme, P., & Newman, M. E. (2006). Nonequilibrium phase transition in the coevolution of networks and opinions. *Physical Review E*, *74*(5), 056108.

Hunter, J. E., Danes, J. E., & Cohen, S. H. (1984). *Mathematical models of attitude change: Change in single attitudes and cognitive structure* (Vol. 1). Academic Press.

Jager, W., & Amblard, F. (2005). Uniformity, bipolarization and pluriformity captured as generic stylized behavior with an agent-based simulation model of attitude change. *Computational & Mathematical Organization Theory*, *10*(4), 295–303.

Jayles, B., Kim, H., Escobedo, R., Cezera, S., Blanchet, A., Kameda, T., Sire, C., & Theraulaz, G. (2017). How social information can improve estimation accuracy in human groups. *Proceedings of the National Academy of Sciences*, *114*(47), 12620–12625.

Kunda, Z. (1990). The case for motivated reasoning. *Psychological Bulletin*, *108*(3), 480.

Li, Y., Fan, J., Wang, Y., & Tan, K.-L. (2018). Influence maximization on social graphs: A survey. *IEEE Transactions on Knowledge and Data Engineering*, *30*(10), 1852–1872.

Lorenz, J., Neumann, M., & Schröder, T. (2021). Individual attitude change and societal dynamics: Computational experiments with psychological theories. *Psychological Review*, *128*(4), 623.

Lorenz, J., Rauhut, H., Schweitzer, F., & Helbing, D. (2011). How social influence can undermine the wisdom of crowd effect. *Proceedings of the National Academy of Sciences*, *108*(22), 9020–9025.

Lu, W., Chen, W., & Lakshmanan, L. V. (2015). From Competition to Complementarity: Comparative Influence Diffusion and Maximization. *Proceedings of the VLDB Endowment*, *9*(2).

Madsen, J. K., Bailey, R. M., & Pilditch, T. D. (2018). Large networks of rational agents form persistent echo chambers. *Scientific Reports*, *8*(1), 1–8.

Nickerson, R. S. (1998). Confirmation Bias: A Ubiquitous Phenomenon in Many Guises. *Review of General Psychology*, *2*(2), 175–220. https://doi.org/10.1037/1089-2680.2.2.175

Osgood, C. E., & Tannenbaum, P. H. (1955). The principle of congruity in the prediction of attitude change. *Psychological Review*, *62*(1), 42.

Perfors, A., & Navarro, D. J. (2019). *Why do echo chambers form? The role of trust, population heterogeneity, and objective truth.* 918–923.

Salzarulo, L. (2006). A continuous opinion dynamics model based on the principle of meta-contrast. *Journal of Artificial Societies and Social Simulation*, *9*(1).

Sherif, M., & Hovland, C. I. (1961). *Social judgment: Assimilation and contrast effects in communication and attitude change.* (pp. xii, 218). Yale Univer. Press.

Sunstein, C. R. (2009). *Going to extremes: How like minds unite and divide*. Oxford University Press.

Takács, K., Flache, A., & Mäs, M. (2016). Discrepancy and disliking do not induce negative opinion shifts. *PloS One*, *11*(6), e0157948.

Wagner, C. (1978). Consensus through respect: A model of rational group decision-making. *Philosophical Studies: An International Journal for Philosophy in the Analytic Tradition*, *34*(4), 335–349.

Weisbuch, G., Deffuant, G., & Amblard, F. (2005). Persuasion dynamics. *Physica A: Statistical Mechanics and Its Applications*, *353*, 555–575.